\pdfoutput=1
\documentclass[cernpreprint, atlasdraft=false, texlive=2025, UKenglish, texmf, orcidlogo]{atlasdoc}
\usepackage[subfigure]{atlaspackage}
\usepackage{atlasbiblatex}

\usepackage{atlasphysics}
\graphicspath{{logos/}{figures/}}

\usepackage{ANA-EXOT-2018-60-PAPER-defs}

\usepackage{diagbox} %
\usepackage{multirow} %

\AtlasTitle{Search for single production of vector-like quarks
decaying into $W(\ell\nu)b$ in $pp$ collisions at $\sqrt{s} = 13$~TeV with the ATLAS detector}

\AtlasAbstract{A search for single production of a vector-like quark
$Q$, which could be either a singlet $T$, with charge $\tfrac23$, or
a $Y$ from a $(T,B,Y)$ triplet, with charge $-\tfrac43$, is
performed using data from proton--proton collisions at a
centre-of-mass energy of 13~TeV. The data correspond to the full
integrated luminosity of 140~fb$^{-1}$ recorded with the ATLAS
detector during Run~2 of the Large Hadron Collider. The analysis
targets $Q \to Wb$ decays where the $W$ boson decays
leptonically. The data are found to be consistent with the expected
Standard Model background, so upper limits are set on the
cross-section times branching ratio, and on the coupling of the $Q$
to the Standard Model sector for these two benchmark models.
Effects of interference with the Standard Model background are taken
into account. For the singlet $T$, the 95\% confidence level limit
on the coupling strength $\kappa$ ranges between~0.22 and~0.52 for
masses from 1150 to 2300~GeV.  For the $(T,B,Y)$ triplet, the limits
on $\kappa$ vary from~0.14 to~0.46 for masses from 1150 to 2600~GeV.
}

\author{The ATLAS Collaboration}

\AtlasRefCode{EXOT-2018-60}

\PreprintIdNumber{CERN-EP-2025-098}

\AtlasHeading{ATLAS Paper}

\AtlasJournal{JHEP}
\AtlasJournalRef{JHEP 12 (2025) 012}
\AtlasDOI{10.1007/JHEP12(2025)012}

\hypersetup{pdftitle={ATLAS document},pdfauthor={The ATLAS Collaboration}}

\begin{document}

\maketitle

\tableofcontents
\clearpage

\section{Introduction}
\label{sec:intro}


Vector-like quarks (VLQs) are hypothetical spin-$\tfrac12$ coloured
particles with left-handed and right-handed components that transform
in the same way under the Standard Model (SM) gauge group. Their
masses, \mQ, are therefore not generated by a Yukawa coupling to the Higgs
field~\cite{Aguilar-Saavedra:2013qpa}.

Although the Higgs-boson production and decay rates observed at the
Large Hadron Collider (LHC)~\cite{HIGG-2012-27,CMS-HIG-12-028} exclude
a fourth generation of perturbatively coupled chiral
quarks~\cite{Eberhardt:2012gv}, the effects of VLQs via quantum-loop
processes~\cite{Aguilar-Saavedra:2013qpa} are much smaller than the
uncertainty in the current measurements, so this family of models
remains compatible with the latest indirect experimental constraints.

VLQs appear in several extensions of the SM addressing the hierarchy
problem, such as extra dimensions~\cite{Randall:1999ee}, composite
Higgs~\cite{Kaplan:1983sm,Vignaroli:2012sf} and Little
Higgs~\cite{Schmaltz:2005ky} models, where they are added to the SM in
multiplets. They can also appear in supersymmetric
models~\cite{Martin:2009bg} and are able to stabilise the electroweak
vacuum~\cite{Xiao:2014kba}. In many such
models, VLQs mix mainly with the SM quarks of the third generation because
of the large masses of the bottom and top
quarks~\cite{delAguila:1982fs,Cacciapaglia:2011fx}.
In addition, the radiative neutrino mass generation mechanism usually
requires vector-like quarks to generate the tiny masses of neutrinos
via loops involving
VLQs~\cite{PhysRevD.85.073005,PhysRevD.89.055013,Batra:2025gzy}.

\begin{figure}[tbhp]
\centering
\includegraphics[width=0.5\textwidth]{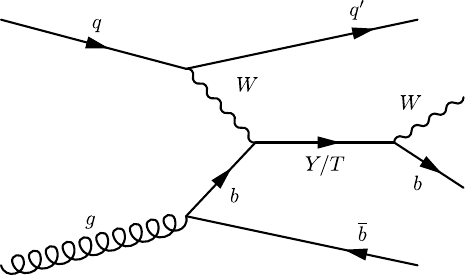}
\caption{Leading-order
Feynman diagram for single $Y$/$T$ production in $Wb$ fusion and
subsequent decay into $Wb$.}
\label{fig:FeynmanDiagram}
\end{figure}

The set of possible VLQs are the $B$ and $T$ quarks, named in analogy with
the SM $b$- and $t$-quarks and with their electromagnetic (EM) charges
$-\tfrac13$ and $+\tfrac23$, and $X$ and $Y$ quarks
with more exotic charges $+\tfrac53$ and $-\tfrac43$.
This paper presents a direct search for a heavy vector-like $T$
quark %
or $Y$ quark, or their antiparticles.\footnote{Antiparticles are
implicitly included in all discussions.} The different charges in the
final states, $Y^{+4/3}\to W^+ \bar{b}^{+1/3}$ and $T^{+2/3} \to W^+
b^{-1/3}$, are not exploited here.
The analysis focuses on the production of a single VLQ ($Q=T$ or $Y$)
produced via $Wb$ fusion ($qg \to q^\prime Q b$), which then decays
back into $Wb$ ($Q\to Wb$). The subsequent leptonic $W$ decays are
considered, complementing a similar search in the fully hadronic final
state~\cite{EXOT-2022-43}. While it is possible for the VLQ to couple
also to light quarks, as e.g. investigated in another recent search in
Ref.~\cite{EXOT-2021-35}, only coupling to $b$-quarks is considered
here.
An example of a leading-order Feynman diagram for this production
process is shown in Figure~\ref{fig:FeynmanDiagram}.

\subsection{VLQ phenomenology}

Several Lagrangian formulations exist for VLQ interactions, including
a renormalisable extension of the SM parameterised by chiral left- and
right-handed mixing angles $\theta_\mathrm{L}$ and $\theta_\mathrm{R}$
related through multiplet symmetries~\cite{Aguilar-Saavedra:2013qpa},
and a non-renormalisable effective theory parameterised by chiral
$\cL$ and $\cR$ couplings~\cite{DeSimone:2012fs,Matsedonskyi:2014mna}.
The principal difference between these approaches is that the
non-renormalisable model's Lagrangian allows for additional terms that
could result in larger production cross-sections. These formalisms
were used in previous ATLAS and CMS
papers~\cite{EXOT-2013-17,EXOT-2014-12,EXOT-2016-35,EXOT-2016-17,CMS-B2G-15-008,CMS-B2G-16-005,CMS-B2G-17-007}, but here the simplified
Lagrangian from Ref.~\cite{Buchkremer:2013bha} is
adopted:
\begin{equation}\label{eqn:SimpLag}
\mathcal{L}
= \sum_{Q, q, \zeta} \left[
\frac{g_w}{\sqrt{2}}{\kappa}_{\!\zeta}^{Qq} \bar{Q}\slashed{W} P_{\!\zeta} q +
\frac{g_w}{2c_W} \tilde{\kappa}_{\!\zeta}^{Qq} \bar{Q}\slashed{Z} P_{\!\zeta} q +
\hat{\kappa}_{\zeta}^{Qq} H \bar{Q} P_{\!\zeta} q\right] + \mathrm{h.c.}
\end{equation}
Here, $Q$ represents the VLQ flavour(s), $\zeta$ is the
chirality and $P_{\!\zeta}$ is its projection operator, $q$
represents the interacting SM quark, $c_W$ and $g_W$ are electroweak parameters, and ${\kappa}_\zeta^{Qq}, \tilde{\kappa}_\zeta^{Qq},$ and $\hat{\kappa}_\zeta^{Qq}$ are the electroweak couplings
between $Q$ and $q$ when mediated by the $W$, $Z$, and $H$ bosons respectively.

This paper focuses on the $Q = \{T, Y\}$ VLQs and their $W$
interaction mode with $q=b$, i.e.~the first term in
Eq.~(\ref{eqn:SimpLag}) and a reduction of the sum to chiral
projections only. Since the relations between chiral states are
determined by the multiplet structure, the results of this analysis
are interpreted in terms of the cross-section times branching ratio
and a single coupling-strength parameter
$\kappa \equiv \kappa_\zeta^{Qb}$ with an assumed branching ratio, as
a function of the VLQ mass.

\mbox{Vector-like $T$ quarks} can belong to any weak-isospin
SU(2) multiplet, while vector-like $Y$ quarks cannot exist as singlets.
This analysis focuses on $Y$ quarks from a
$(T,B,Y)$ triplet, or a singlet $T$ quark. $T$
quarks in a $(T,B,Y)$ triplet do not couple to
$Wb$~\cite{Aguilar-Saavedra:2013qpa}.
For singlet $T$ quarks, the branching ratios ($\mathcal{B}$) are
model- and mass-dependent, but in the high-mass limit considered in
this analysis, they converge towards $(Wb:Zt:Ht) =
(2:1:1)$~\cite{Aguilar-Saavedra:2013qpa}.
Due to its $-\tfrac43$ charge, $Y$ quarks can be produced singly in $pp$
collisions only via $Wb$ fusion, and they can decay only into $Wb$.
Consequently, $\BR{Y \to Wb}=100\%$.

In this paper it is assumed that $T$ quarks are
produced in $Wb$ fusion only.
For single production of a $T$ quark, $Zt$ fusion could in principle
contribute as well, but is neglected in this $T$-singlet search primarily
because the cross-section for $Zt$ fusion is much smaller than for
$Wb$ fusion. For equal values of the $TZt$ and $TWb$ couplings, it
is about one order of magnitude smaller~\cite{Matsedonskyi:2014mna}.
However, for the $T$-singlet case, the $TZt$ coupling is about a factor of
$\sqrt{2}$ smaller than the $TWb$ coupling. Since the single-VLQ
production cross-section scales with the coupling squared, the
cross-section for $Zt$ fusion is about an extra factor
of two lower than that for $Wb$ fusion. Similarly, $\BR{T \to Zt}$
is about a factor of two smaller than $\BR{T \to Wb}$.
In addition, the selection efficiency for $tZ \to T \to Wb$ events in
this search is about a factor of two smaller than for
$bW \to T \to Wb$, because in $tZ \to T \to Wb$ the accompanying top
quark from the gluon splitting leads to additional jets in the final
state (see Section~\ref{sec:pre-selection}).

Single production of vector-like quarks is enabled by their mixing
with the SM quarks, whereas pair production is dominated by the
model-independent QCD coupling.
As a result, searches for VLQ pair production can set absolute bounds
on the VLQ masses, while searches for single production provide a
complementary mapping of their couplings to the SM
quarks as a function of those masses.
In addition, single production can become the dominant VLQ production
mechanism at the LHC for high VLQ masses (about 1000~\GeV\ for a
coupling of 0.5), where the energy threshold for pair production
suppresses that cross-section enough to offset the difference between
the coupling strengths of QCD and weak interactions.

\subsection{Previous studies}

The ATLAS and CMS Collaborations have published searches for single
and pair production of vector-like $T$ quarks in all three decay
channels~\cite{EXOT-2013-18, EXOT-2016-15,EXOT-2016-17,
CMS-B2G-14-002,CMS-B2G-12-015, CMS-B2G-17-003, EXOT-2013-16,
EXOT-2014-12, EXOT-2016-13, CMS-B2G-16-024, CMS-B2G-16-001,
EXOT-2016-14, CMS-B2G-17-011, EXOT-2014-12, CMS-B2G-15-008,
CMS-B2G-16-006, CMS-B2G-17-007, EXOT-2013-17, CMS-B2G-16-005,
EXOT-2016-35, EXOT-2017-17}, and have set 95\% confidence level (CL)
lower limits on the $T$- and $Y$-quark masses.
Assuming $\br=100\%$ for the corresponding decay channel, the
strongest observed $T$-quark mass limits are $m_T > 1430$~\GeV\ for $T
\to Ht$~\cite{EXOT-2016-13}, 1340~\GeV\ for $T \to
Zt$~\cite{EXOT-2016-35} and 1350~\GeV\ for $T \to
Wb$~\cite{CMS-B2G-17-003}, independent of the size of the $c^{Wb}$
couplings.
Reinterpretation studies combining LHC measurements~\cite{Buckley:2020wzk}
observe a $\kappa$-dependent sensitivity to single-production at higher
masses, setting $m_{T/Y}$ limits between 1500 to 2000~\GeV for high
values of $\kappa$ (typically $\kappa > 0.6$) depending on the assumed
multiplet structure.
In Ref.~\cite{EXOT-2017-17}, seven individual ATLAS analyses searching for
$B\bar{B}$ or $T\bar{T}$ pair production were combined, significantly
improving on the model-independent cross-section limits from individual
analyses and, in particular, excluding $T$-quark masses lower than
1310~\GeV\ for any combination of decays into SM particles.
The observed lower limit on the pair-produced $Y$-quark's mass is
1350~\GeV~\cite{EXOT-2016-14}.

These searches also reported limits as a function of the assumed
branching ratios.
The strongest observed limits are $m_T > 1310$~\GeV\ and $m_T >
1280$~\GeV\ for a weak-isospin doublet~\cite{EXOT-2016-13} and
singlet~\cite{CMS-B2G-17-011} respectively.
In the $T \to Zt$ decay channel, assuming a mixing parameter $\sin
\theta_{\mathrm{L}}$ as low as $0.7$, $T$ quarks with masses between
$450~\GeV$ and $650~\GeV$ are excluded~\cite{EXOT-2013-17}, while for
$QWb$ couplings satisfying $\cTwoL + \cTwoR = 1$ and assuming
$\BR{T\to Wb} = 50\%$, the observed lower limit on the $T$-quark mass
is $950~\GeV$~\cite{EXOT-2014-12,EXOT-2016-17}.
The CMS Collaboration studied single $T$- and $Y$-quark production
using various subsets of \RunTwo data at $\sqrt{s} = 13~\TeV$ collected
in 2015~\cite{CMS-B2G-16-005,CMS-B2G-15-008, CMS-B2G-16-001,
CMS-B2G-16-006, CMS-B2G-17-007}, and set upper limits on the
single-$T$-quark production cross-section times $\BR{T\to Ht}$ that
are between $\mathrm{0.31~pb}$ and $\mathrm{0.93~pb}$ for $T$-quark
masses in the range 1000--1800~\GeV~\cite{CMS-B2G-16-005}.
Single-$T$-quark production cross-section times $\BR{T \to Zt}$ upper
limits between $\mathrm{0.98~pb}$ and 0.15~pb ($\mathrm{0.6~pb}$ and
$\mathrm{0.13~pb}$) were set for $T$-quark masses in the range 700--1700~\GeV\ in
the left-handed $T(b)$ (right-handed $T(t)$) production channel~\cite{CMS-B2G-16-001}.
For a mass of 1000~\GeV, a $T$-quark production cross-section times
branching ratio above 0.8~pb (0.7~pb) is excluded for the $T \to Ht$
decay channel assuming left-handed (right-handed) coupling of the $T$
quark to SM particles~\cite{CMS-B2G-15-008}.
For $Y$ quarks with a coupling of $0.5$ and $\BR{Y \to Wb} = 100\%$,
the observed and expected lower limits on the VLQ mass are $m_Y >
1400$ and~$1000~\GeV$ respectively~\cite{CMS-B2G-16-006}.

\subsection{Analysis overview}

This paper describes a search for $Q \to Wb$ ($Q=T$ or $Y$)
production, with the $W$ boson decaying via $W\to \ell \nu$, with
$\ell = e$ or $\mu$. It uses the full  \RunTwo dataset from $\sqrt{s}=\SI{13}{\TeV}$
\pp{} collisions recorded by the ATLAS detector at the LHC between
2015 and 2018, corresponding to an integrated luminosity of
140~\ifb~\cite{DAPR-2021-01,LUCID2}.
A set of single-electron~\cite{TRIG-2018-05} and single-muon
triggers~\cite{TRIG-2018-01} were used, with requirements
depending on the lepton flavour and data-taking period.
All detector subsystems were required to be operational during data
taking and data-quality requirements~\cite{DAPR-2018-01} were applied.

The analysis in this paper supersedes the previous ATLAS analysis~\cite{EXOT-2016-17}
using \RunTwo data with an integrated luminosity of 36.1~\ifb
recorded in 2015--2016.
A singly produced $Q$, followed by a leptonically decaying $W$ boson
results in a lepton-plus-jets signature characterised by the presence of
exactly one electron or muon,\footnote{Electrons and muons from decays
of $\tau$-leptons from $W \to \tau \nu$ might also pass this
selection.} three or more jets and missing transverse momentum from
the escaping neutrino.
The analysis is optimised to search for decays of massive VLQs with a
high-momentum $b$-jet (i.e. coming from the hadronisation of a
$b$-quark) in the final state.
The $b$-jet and the charged lepton originating from the $Q$ decay are
approximately back-to-back in the transverse plane since both
originate from the decay of a heavy object.

The outgoing light quark in the process depicted in
Figure~\ref{fig:FeynmanDiagram} often produces a jet in the forward
region of the detector.
The second $b$-jet from the gluon splitting may be observed in either
the forward or central region of the detector.
Since this $b$-jet typically has low energy, it often falls outside
the detector acceptance.

The main background process with a prompt lepton arises from
$W$-boson production in association with jets (\Wjets), followed by
top-quark pair ($t\bar{t}$) production and single top-quark
production, with smaller background contributions from $Z$-boson
production in association with jets (\Zjets), and diboson production ($WW$,
$WZ$, and $ZZ$).
Multijet events also contribute to the selected sample via the
misidentification of a jet or a photon as an electron or the presence
of a non-prompt electron or muon.
To estimate the backgrounds from \ttbar and \Wjets events in a
consistent and robust fashion, several control regions and validation
regions are used. These are defined to be orthogonal to the signal region
in order to provide independent data samples enriched in particular
background sources.
The reconstructed mass of the heavy VLQ candidate is used as the
discriminating variable in a binned likelihood fit to test for the
presence of a signal, taking into account interference with SM
background processes.
%


%

%
\section{ATLAS detector}
\label{sec:detector}

%
\newcommand{\AtlasCoordFootnote}{%
ATLAS uses a right-handed coordinate system with its origin at the nominal interaction point (IP)
in the centre of the detector and the \(z\)-axis along the beam pipe.
The \(x\)-axis points from the IP to the centre of the LHC ring,
and the \(y\)-axis points upwards.
Polar coordinates \((r,\phi)\) are used in the transverse plane,
\(\phi\) being the azimuthal angle around the \(z\)-axis.
The pseudorapidity is defined in terms of the polar angle \(\theta\) as \(\eta = -\ln \tan(\theta/2)\) and is equal to the rapidity
$ y = (1/2) \ln \left[ (E + p_z)/(E - p_z) \right] $ in the relativistic limit.
Angular distance is measured in units of \(\Delta R \equiv \sqrt{(\Delta y)^{2} + (\Delta\phi)^{2}}\).}

\label{sec:atlas2b}

The ATLAS detector~\cite{PERF-2007-01} at the LHC~\cite{Evans:2008zzb} covers nearly the entire solid angle around the collision point.\footnote{\AtlasCoordFootnote}
It consists of an inner tracking detector (ID) surrounded by a thin superconducting solenoid, electromagnetic and hadronic calorimeters,
and a muon spectrometer incorporating three large superconducting air-core toroidal magnets.

The inner-detector system (ID) is immersed in a \qty{2}{\tesla} axial magnetic field
and provides charged-particle tracking in the range \(|\eta| < 2.5\).
The high-granularity silicon pixel detector covers the vertex region and typically provides four measurements per track,
the first hit generally being in the insertable B-layer (IBL) installed before Run~2~\cite{ATLAS-TDR-19,PIX-2018-001}.
It is followed by the SemiConductor Tracker (SCT), which usually provides eight measurements per track.
These silicon detectors are complemented by the transition radiation tracker (TRT),
which enables radially extended track reconstruction up to \(|\eta| = 2.0\).
The TRT also provides electron identification information
based on the fraction of hits (typically 30 in total) above a higher energy-deposit threshold corresponding to transition radiation.

The calorimeter system covers the pseudorapidity range \(|\eta| < 4.9\).
Within the region \(|\eta|< 3.2\), electromagnetic (EM) calorimetry is provided by barrel and
endcap high-granularity lead/liquid-argon (LAr) calorimeters,
with an additional thin LAr presampler covering \(|\eta| < 1.8\)
to correct for energy loss in material upstream of the calorimeters.
Hadronic calorimetry is provided by the steel/scintillator-tile calorimeter,
segmented into three barrel structures within \(|\eta| < 1.7\), and two copper/LAr hadronic endcap calorimeters.
The solid angle coverage is completed with forward copper/LAr and tungsten/LAr calorimeter modules
optimised for electromagnetic and hadronic energy measurements, respectively.

The muon spectrometer (MS) comprises separate trigger and
high-precision tracking chambers measuring the deflection of muons in a magnetic field generated by the superconducting air-core toroidal magnets.
The field integral of the toroids ranges between \num{2.0} and \qty{6.0}{\tesla\metre}
across most of the detector.
Three layers of precision chambers, each consisting of layers of monitored drift tubes, cover the region \(|\eta| < 2.7\),
complemented by cathode-strip chambers in the forward region, where the background is highest.
The muon trigger system covers the range \(|\eta| < 2.4\) with resistive-plate chambers in the barrel, and thin-gap chambers in the endcap regions.

The luminosity is measured mainly by the LUCID--2~\cite{LUCID2} detector that records Cherenkov light produced in the quartz windows of photomultipliers located close to the beampipe.

Events were selected by the first-level trigger system implemented in custom hardware,
followed by selections made by algorithms implemented in software in the high-level trigger~\cite{TRIG-2016-01}.
The first-level trigger accepted events from the \qty{40}{\MHz} bunch crossings at a rate close to \qty{100}{\kHz},
which the high-level trigger further reduced in order to record complete events to disk at about 1.25~\unit{\kHz}.

A software suite~\cite{SOFT-2022-02} is used in data simulation, in the reconstruction
and analysis of real and simulated data, in detector operations, and in the trigger and data acquisition
systems of the experiment.


%

%
%

%
\section{Background and signal simulation}
\label{sec:samples}


Except for the multijet background from non-prompt and fake leptons,
the expected background and signal event contributions are estimated using
Monte Carlo (MC) simulation samples, processed either through the full
ATLAS detector simulation~\cite{SOFT-2010-01} based on
\GEANT~\cite{Agostinelli:2002hh} or through a faster
simulation making use of parameterised showers in the
calorimeters.
Inelastic proton-proton interactions, modelled with
\PYTHIA[8.186]~\cite{Sjostrand:2007gs}
and fully simulated, were overlayed in the fast or full detector
simulation, according to the luminosity profile of the recorded data,
to simulate activity from additional $pp$ collisions in the same and
nearby bunch crossings (in-time and out-of-time pile-up.)

All simulated events were processed using the same reconstruction
algorithms and analysis selection requirements as for the data. Small
corrections, obtained from comparisons of simulated events with data
in dedicated control regions, were applied to trigger and
object-reconstruction efficiencies as well as detector resolutions, to
better model the observed response.

The main configurations of the MC samples used in this search are
summarised in Table~\ref{table:MCsamples}. All simulated event
samples, other than for processes modelled using the {\SHERPA}
generator~\cite{Gleisberg:2008ta}, use
\EVTGEN~\cite{Lange:2001uf} to model the decays of
heavy-flavour hadrons.

\begin{table}[tbhp]
\centering
\caption{\label{table:MCsamples} MC generator configurations used to
model the signal and background processes. All \Pythia generation
used the A14 tune for the parton shower, hadronisation and underlying
event, together with the \NNPDF[2.3lo] PDF set; all \Sherpa
generation used the default \Sherpa tune for its generator
version. The PDF set used for hard-process modelling, and the
highest order of the perturbative QCD (pQCD) calculation used for
cross-section normalisation are given for each sample. All
background processes
were generated at NLO in QCD. The signal processes were generated at
LO, but normalised to cross-sections calculated at NLO in QCD.
The LO cross-sections calculated for the $Yqb$ signal processes in
the simulation were normalised to the NLO theoretical
cross-section taken from Ref.~\cite{Cacciapaglia:2018qep}.}
\small
\begin{tabular}{lllll}
\toprule

Process & Generator & Hard-process & Inclusive cross-section \\
& + parton showering/hadronisation & PDF set & order in pQCD \\
\midrule
\textbf{$Tqb$ signal} & \MGNLO[2.6.5] & \NNPDF[2.3] & NLO \\
($Yqb$ by reweighting) & + \PYTHIA[8.210]& & \\
\addlinespace
\textbf{\ttbar} & \POWHEGBOX[v2] & \NNPDF[3.0] & NNLO \\
& + \PYTHIA[8.230] & & & \\
\addlinespace
\textbf{Single top} & \POWHEGBOX[v2]  & \NNPDF[3.0] & NNLO \\
& + \PYTHIA[8.230] & & &  \\
\addlinespace
\textbf{Dibosons}   & \SHERPA[2.2.2]  & \NNPDF[3.0] & NLO \\
\textbf{$WW$, $WZ$, $ZZ$} & & & \\
\addlinespace
\textbf{\Zjets} & \SHERPA[2.2.1]  & \NNPDF[3.0] & NNLO \\
\textbf{\Wjets} & \SHERPA[2.2.11] & \NNPDF[3.0] & NNLO \\
\addlinespace
\textbf{$\ttbar V$} & \MGNLO[2.2.3]  & \NNPDF[2.3] & NLO \\
& + \PYTHIA[8.210]& & & \\
\addlinespace
\textbf{$\ttbar H$} & \POWHEGBOX[v2]  & \NNPDF[3.0] & NLO \\
& + \PYTHIA[8.230]& & &  \\
\bottomrule
\end{tabular}
\end{table}

\subsection{Simulated signal samples}
\label{sec:signal-samples}

Simulated events for $T$-quark signal processes were generated at
leading order (LO) in QCD in the four-flavour scheme with the
\MGNLO[2.6.5] generator~\cite{Alwall:2014hca} using the
\NNPDF[2.3nlo] PDF set~\cite{Ball:2012cx}, interfaced
to \Pythia[8]~\cite{Sjostrand:2014zea} for parton-showering and
hadronisation using the A14 set of tuned parameters with
the \NNPDF[2.3lo] PDFset~\cite{ATL-PHYS-PUB-2014-021}. The VLQ model
from Ref.~\cite{Buchkremer:2013bha} was used in the computation of the
matrix elements, and all tree-level processes were included. Signal
events were processed with fast simulation.

Other possible decay modes of the $T$ quark ($T \to Zt$ and $T \to
Ht$) have negligible acceptance in this search, confirmed by testing
samples generated for other analyses focused on those channels. The
kinematics of the final-state particles are very similar for
left-handed and right-handed couplings, and hence the acceptances for
the two chiralities were also found to be equal. The LO cross-sections
calculated for the signal processes in the simulation were normalised
to the next-to-leading-order (NLO) benchmark calculation from
Ref.~\cite{Cacciapaglia:2018qep}, which was performed in the
narrow-width approximation, and a correction factor was applied to
account for finite-width effects~\cite{Roy:2020fqf,Roy:2022wjj}.
The validity of these correction factors is limited to VLQ relative widths
satisfying $\Gamma_Q/\mQ < 0.5$, following Ref.~\cite{Deandrea:2021vje}.

Separate event samples were produced for $T$-quark masses from
1100~\GeV{} to 2700~\GeV{} in steps of 200~\GeV, with the coupling
between the $T$ quark and the gauge boson set to
$\kappa=1.0$.\footnote{This coupling $\kappa$ corresponds to the
parameter $\kappa_T$ in Ref.~\cite{Buchkremer:2013bha}. It is also
related to the $\cLR{}$ coupling parameters in
Ref.~\cite{Matsedonskyi:2014mna} via $\kappa = f(\mQ) \,
\cLR{}/\sqrt{2}$, where $f^2(\mQ) \approx 1+\mathcal{O}(\mQ^{-4})$ and
\mQ is the VLQ mass in \GeV.}
In order to determine the signal yields and acceptances for VLQ masses
and couplings different from those at the 200~\GeV mass steps and the nominal
coupling strength of $\kappa = 1.0$, event-by-event reweighting
factors were applied, following the same procedure as used in
Ref.~\cite{EXOT-2016-17}.
Since the kinematic distributions of the decay products for the $T$
quark and $Y$ quark in the $Wb$ decay channel are the same, these
$T$-quark signal samples were also used to derive the results for the
$Yqb$ signals, via further event reweighting.

Interference effects between the amplitudes for VLQ signal production
and the SM are possible, and depend on the VLQ multiplet model.  In
this analysis, two
scenarios are considered:
\begin{enumerate}
\item $T$-quark production in a $T$-singlet model if the $T$ quark
has only a left-handed coupling~\cite{Aguilar-Saavedra:2013qpa},
which could interfere with SM $t$-channel single-top-quark
production if the top quark is far off-shell.
\item $Y$-quark production with only a left-handed coupling, which
could interfere with SM electroweak $W^{-}bq$
production.\footnote{The charge-conjugated state $W^{+}\bar{b}q$
interferes with the $\bar{Y}$ quark.}  This scenario is realised
in a $(T,B,Y)$ triplet model for which the right-handed coupling
is heavily suppressed~\cite{Aguilar-Saavedra:2013qpa}. Since the
$T$ does not couple to $Wb$ in the $(T,B,Y)$ triplet model,
$T$-quark production does not contribute to the final state under
consideration.
\end{enumerate}

To account for the effects of interference, further reweighting
factors were calculated for each mass point and each
$\kappa$~\cite{EXOT-2016-17}, and for each interference scenario ($T$
or $Y$ as described above), using
\MGNLO[2.6.5]. The squared matrix element for the process $pp \to Wbq$
is given by
\begin{linenomath*}
\begin{equation*}
|M|^2 = |M_{\mathrm{SM}}|^2 + |M_{\mathrm{VLQ}}|^2 + 2 \mathfrak{Re}\!\left( M_{\mathrm{SM}}^{*} M_{\mathrm{VLQ}} \right) \, ,
\end{equation*}
\end{linenomath*}
and hence the total cross-section for $pp \to Wb q$ at LO can be
written as $\sigma_{\mathrm{tot}}^{\mathrm{LO}} =
\sigma_{\mathrm{SM}}^{\mathrm{LO}} +
\sigma_{\mathrm{VLQ}}^{\mathrm{LO}} +
\sigma_{\mathrm{I}}^{\mathrm{LO}}$
where $\sigma_{\mathrm{SM}}^{\mathrm{LO}}$ is the LO SM cross-section,
$\sigma_{\mathrm{VLQ}}^{\mathrm{LO}}$ is the LO VLQ cross-section,
and $\sigma_{\mathrm{I}}^{\mathrm{LO}}$ is the contribution from interference effects.
Therefore, in the final statistical analysis, the interference
contribution $\sigma_{\mathrm{I}}^{\mathrm{LO}}$ is related to the
total signal cross-section, scaling as $\sqrt{\sigma_{\mathrm{VLQ}}^{\mathrm{LO}}}$.

\subsection{Simulated background samples}
\label{sec:background-samples}

The main backgrounds originate from SM $W$-boson production in
association with jets (\Wjets) and from \ttbar{} and single-top-quark
production, with smaller contributions from $Z$-boson production in
association with jets ($Z$+jets) and from diboson ($WW$, $WZ$, $ZZ$)
production. Samples for all SM background processes were simulated
with the full \GEANT model of the ATLAS detector. This
section summarises the different generators used for each process,
and their nominal configurations; the details of
systematic-uncertainty estimation for these theoretical predictions are
discussed in Section~\ref{sec:systematics}.

Standard Model \ttbar{} events were produced with the NLO generator
\POWHEGBOX[v2]~\cite{Nason:2004rx,Frixione:2007vw,Alioli:2010xd} with
the \NNPDF[3.0nlo]~\cite{Ball:2014uwa} PDF set in the matrix-element
calculations. Parton-showering, hadronisation and the underlying event
were simulated using \Pythia[8.230] with the \NNPDF[2.3lo] PDF
set and the A14 set of tuned
parameters~\cite{ATL-PHYS-PUB-2014-021}. This sample was normalised to
the integrated cross-section calculated by
\TOPpp[2.0]~\cite{Cacciari:2011hy,Beneke:2011mq,Baernreuther:2012ws,Czakon:2012zr,Czakon:2012pz,Czakon:2013goa,Czakon:2011xx}
at next-to-next-to-leading order (NNLO) in QCD, including resummation
of next-to-next-to-leading logarithmic soft-gluon terms. Corrections for
NLO electroweak effects in \ttbar events~\cite{Kuhn:2013zoa} were
applied to the generated top-quark kinematics as a function of both the
flavour and centre-of-mass energy of the initial partons and the decay
angle of the top quarks in the centre-of-mass frame of the initial partons.
The \hdamp parameter in \POWHEGBOX, which controls the matching
of the matrix element to the parton shower and effectively regulates
the degree of high-\pt\ radiation, was set to
$1.5m_t$~\cite{ATL-PHYS-PUB-2017-007}, where $m_t = \SI{172.5}{\GeV}$.
Two alternative \ttbar samples with matrix elements calculated by \AMCatNLO
and \POWHEGBOX respectively, and using \Herwig[7.1.3]~\cite{Bahr:2008pv}
instead of  \Pythia[8.230] for showering and hadronisation, were used to
estimate the effects of using different matching prescriptions.

Single-top-quark background event samples corresponding to the $Wt$,
$t$-channel and $s$-channel production mechanisms were generated with
\POWHEGBOX[v2] with the \NNPDF[3.0nlo] PDF set, and interfaced to
\Pythia[8.230], which used the A14 tune and
\NNPDF[2.3lo] PDF set.  Overlaps between the \ttbar and $Wt$
final states were removed using the \enquote{diagram removal} scheme~\cite{Frixione:2005vw, Frixione:2008yi}.
Another set of single-top-quark samples were generated using \MGNLO
interfaced to \Pythia[8.230] to determine the systematic
uncertainties associated with the generation of the hard process.
Additional single-top-quark samples were generated using the same
\POWHEGBOX{} settings as the nominal sample, but with \Herwig[7.0.4]
for parton-showering, hadronisation, and the underlying event.

Samples of \Wjets and \Zjets events were generated using
\Sherpa[2.2.11] and \Sherpa[2.2.1],
respectively~\cite{Gleisberg:2008ta,Hoeche:2009rj}. The former
includes improvements to the statistical population of
phase-space regions where the \Wjets production cross-section is low.
The matrix-element calculation was performed with up to two partons at
NLO and up to four partons at leading order using
\textsc{Comix}~\cite{Gleisberg:2008fv} and
\OPENLOOPS~\cite{Cascioli:2011va}. The matrix-element
calculation, performed using the \NNPDF[3.0nnlo] PDF
set~\cite{Ball:2014uwa}, was merged with the {\Sherpa} parton
shower~\cite{Schumann:2007mg} using the \MEPSatNLO
prescription~\cite{Hoeche:2012yf}.
Alternative \Wjets and \Zjets event samples were produced with \MGNLO[2.2.1], using the
\NNPDF[3.0nlo] PDF set, and interfaced to \Pythia[8.186], which used the
A14 tune and ~\NNPDF[2.3lo] PDF set~\cite{ATL-PHYS-PUB-2014-021}.

Diboson events ($WW$/$WZ$/$ZZ$), with one of the bosons decaying
hadronically and the other leptonically, were generated at NLO with
\Sherpa[2.2.2], using the \NNPDF[3.0nnlo] PDF set. Processes
containing up to four electroweak vertices were included. The matrix element
included up to one ($ZZ$) or no ($WW$, $WZ$) additional partons at
NLO and up to three partons at LO, using the same procedure as for
$W/Z$+jets.  All diboson samples were normalised to their NLO
theoretical cross-sections provided by \Sherpa.

Smaller backgrounds were also considered, including ones from
SM \ttbar production with an associated boson: $\ttbar V$ and $\ttbar H$
with $V = W,Z$. The $\ttbar V$ event samples were simulated with \MGNLO
using the \NNPDF[3.0nlo] PDF set and interfaced to
\Pythia[8]~\cite{Sjostrand:2014zea}, which used the A14 tune and
\NNPDF[2.3lo] PDF set. The $\ttbar H$
events were modelled using NLO \POWHEGBOX[v2] and interfaced to
\Pythia[8.230].


%

%
\section{Object definitions}
\label{sec:object-definitions}

This analysis seeks to identify events containing a single VLQ that
decays to a final state with an electron or muon, missing transverse
momentum from a neutrino, and a $b$-jet.

\subsection{Lepton selection}

Electron candidates are reconstructed from isolated energy deposits
(clusters) in the EM calorimeter, each matched to a reconstructed ID
track, within the fiducial region of $ | \eta_{\text{cluster}} | <
2.47 $, where $\eta_{\text{cluster}}$ is the pseudorapidity of the
centroid of the calorimeter energy deposit associated with the
electron candidate. A veto is placed on electrons in the transition
region between the barrel and endcap electromagnetic calorimeters,
$1.37 < |\eta_{\text{cluster}}| < 1.52$.  Electrons are required to
satisfy the \textit{tight-likelihood} identification
criterion~\cite{EGAM-2018-01}, based on shower-shape and
track--cluster matching variables, and must have transverse energy
$\ET = E_\mathrm{cluster}/\cosh(\eta_\mathrm{track})> \SI{27}{\GeV}$,
where $E_\mathrm{cluster}$ is the electromagnetic cluster energy and
$\eta_\mathrm{track}$ the track pseudorapidity.

Muons are reconstructed~\cite{MUON-2018-03} by combining a track
reconstructed in the ID with one in the MS, using the complete track
information from both detectors and accounting for the effects of
energy loss and multiple scattering in the material of the detector
structure. The muon candidates must satisfy the \textit{medium}
selection criteria~\cite{MUON-2018-03} and are required to have a
transverse momentum of at least $27\,\GeV$ and to be in the region
$|\eta|<2.5$.

To reduce the contribution of leptons from hadronic decays (non-prompt
leptons), electrons and muons must satisfy isolation criteria that
apply to both track and calorimeter information, and are tuned to give
an overall efficiency of 98\%, independent of the lepton's $\pT$.
Electron and muon candidates are required to have no
additional tracks within a cone of angular radius
${\Delta}R \equiv \sqrt{(\Delta\eta)^{2} + (\Delta\phi)^{2}}$ around their
directions, with ${\Delta}R = \min\{0.2, 10~\GeV{}\text{/} \pt^e\}$
for electrons and ${\Delta}R = \min\{0.3, 10~\GeV{}\text{/} \pt^\mu\}$ for
muons~\cite{MUON-2018-03}. The significance of the transverse impact
parameter $d_{0}$, calculated relative to the measured beam-line
position, is required to satisfy $|d_{0}/\sigma(d_{0})|< 3 (5)$ for
muons (electrons), where $\sigma(d_{0})$ is the uncertainty in
$d_{0}$. Finally, the lepton tracks are matched to the primary vertex
of the event by requiring the longitudinal impact parameter $z_{0}$ to
satisfy $|z_{0}\sin\theta_{\mathrm{track}}|< 0.5$~mm, where
$\theta_{\mathrm{track}}$ is the polar angle of the
track.\footnote{The longitudinal impact parameter $z_{0}$ is the
difference between the longitudinal position of the track along the
beam line at the point where the transverse impact parameter
($d_{0}$) is measured and the longitudinal position of the primary
vertex.}

\subsection{Jet selection}

Hadronic jets are reconstructed from three-dimensional topological
calorimeter energy clusters and information about the tracks of charged
particles passing through the ID.
Jets are reconstructed with the anti-$\mathit{k_{t}}$
algorithm~\cite{Cacciari:2008gp,Fastjet} with a distance parameter of
$R=0.4$~\cite{JETM-2018-05} using particle flow~\cite{PERF-2015-09}.
Data-quality criteria are imposed to identify jets arising from
non-collision sources or detector noise, and any event containing at
least one such jet is removed~\cite{ATLAS-CONF-2015-029}.  Finally,
jets considered in this analysis are required to have a $\pT >
25~\GeV$. The pseudorapidity acceptance for jets differs between
different selections: central jets are required to have $|\eta| <
2.5$, while forward jets are defined to have $2.5 < |\eta|<
4.5$. Furthermore, jets with a $\pT < 60~\GeV$ and $|\eta| < 2.4$ are
required to satisfy criteria implemented in the jet vertex tagger
(JVT) algorithm~\cite{PERF-2014-03,ATL-PHYS-PUB-2019-026}
designed to
select jets that originate from the hard scattering and reduce the
effect of in-time pile-up. A high value of the JVT corresponds to a
high probability that the jet originated from the primary vertex. Hence,
a requirement on JVT removes contamination from pile-up vertices.

\subsection{Jet flavour-tagging and selection}

The identification of jets from $b$-hadron decays ($b$-tagging) is
fundamental in this analysis.  Jets containing $b$-hadrons are
identified (\enquote{tagged}) by a multivariate discriminant that
combines information about the impact parameters of ID tracks
associated with the jet, the presence of displaced secondary vertices,
and the reconstructed flight paths of $b$- and $c$-hadrons inside the
jet~\cite{FTAG-2019-07}. Jets are considered to be $b$-tagged if the
value of the discriminant, based on the
DL1r~\cite{ATL-PHYS-PUB-2017-003} deep neural-network algorithm, is
larger than a certain threshold.  The threshold criterion, with a
fixed value of 0.46, is applied only to central jets ($|\eta|< 2.5$)
with $\pt > \SI{25}{\GeV}$ and has an efficiency of approximately 85\%
for $b$-jets in simulated \ttbar\ events.  The rejection factor for
jets originating from light quarks and gluons (henceforth referred to
as light-flavour jets) is about 100, and that for jets originating
from charm quarks ($c$-jets) is about 5, determined in simulated
\ttbar\ events. Correction factors are defined to correct the tagging
rates in the simulation to match the efficiencies measured in the data
control samples~\cite{FTAG-2018-01,FTAG-2020-08,FTAG-2019-02}.

\subsection{Overlap removal}

To avoid interpreting the same detector measurement as two or more
objects, an overlap-removal procedure is used to resolve ambiguities.
First, any muon that leaves an energy deposit in the calorimeter and
shares an ID track with an electron is removed.
Then, any electron sharing an ID track with a remaining
muon is removed.
Jets overlapping with identified electron
candidates within a cone of $\Delta R = 0.2$ are removed, as the
jet and the electron are very likely to be the same physics object.
If the nearest jet surviving this requirement is within $\Delta R = 0.4$
of an electron, the electron is discarded; this ensures that remaining
electrons are sufficiently separated from nearby jet activity.
Muons are removed if they are
separated from the nearest jet by $\Delta R < \min\{0.4, 0.04 +
10~\GeV/\pt^{\mu}\}$, where $\pt^{\mu}$ is the muon transverse
momentum, to reduce the background from muons from heavy-flavour
hadron decays inside jets.  Any jet with at least three associated
tracks (within a $\Delta R = 0.2$ cone) around a muon
is kept and the muon removed.
If, instead, the jet has fewer than
three such associated tracks, the muon is kept and the jet is removed;
this avoids an inefficiency for high-energy muons that
suffered significant energy loss in the calorimeter.

\subsection{\MET definition and $W$-boson reconstruction}
\label{sec:met_calc}

The missing transverse momentum \vecMET
(with magnitude \MET) is a measure of the momentum of the particles
that remain undetected. It is defined as the negative vector sum of
the transverse momenta of all selected and calibrated objects in the
event
including a term to account for energy from soft particles
which are not associated with any of the selected
objects. This soft term is calculated from
ID tracks matched to the selected primary vertex, reducing
its dependence on pile-up interactions~\cite{JETM-2020-03}.

The \vecMET is then used as the transverse momentum of the neutrino
from the $W$-boson candidate. The $W$-boson candidate is reconstructed
by summing the four-momenta of the charged lepton and the neutrino.
To obtain the $z$-component of the neutrino momentum ($p_{z,\nu}$),
the invariant mass of the lepton--neutrino system is set to the
$W$-boson mass and the resulting quadratic equation is solved. If no
real solution exists, the magnitude of \vecMET
is varied by the minimum amount required to produce exactly one real
solution to the quadratic equation.  If two real solutions are found,
the one with the smaller $|p_{z,\nu}|$ is used.

\subsection{Reconstruction of the VLQ candidate}

A VLQ candidate for each event is reconstructed from the highest transverse momentum (leading \pt)
$b$-tagged jet
and the decay products of the leptonically decaying $W$-boson candidate, i.e.\
the neutrino (see Section~\ref{sec:met_calc}) and the lepton.
The invariant mass of the
reconstructed VLQ candidate, $\mQreco$, is used in the final
statistical analysis as the main variable that discriminates between
signal and background events.


%

%
\section{Preselection}
\label{sec:pre-selection}

To be considered for the analysis, an event must contain exactly one
electron or muon with $\pT{} > 27$~\GeV and it must
be matched to the corresponding electron or muon trigger object.  The
muon trigger covers the region $|\eta|<2.4$, and its efficiency
decreases at larger $|\eta|$ values.  Events must have at least two
jets with $\pT > 25~\GeV$.  Furthermore, the highest-\pT{} jet
(\enquote{leading jet}) must be a $b$-tagged central jet
($|\eta|<2.5$) with $\pT > 200~\GeV$. The missing transverse momentum
is required to satisfy $\MET{} > 120~\GeV$ and to be separated from
the leading jet by $|\Delta\phi$(leading jet, \vecMET)$| \geq 2$.  The
requirements on the missing transverse momentum reduce the fraction of
events that are diboson events or contain a non-prompt or
misidentified lepton.  These baseline requirements are referred to as
the
\enquote{preselection}.  A summary of the main preselection requirements is
provided in Table~\ref{tab:Preselection}.

\begin{table}[tbp]
\centering
\caption{Summary of the preselection requirements for all signal, control, and validation regions.}
\label{tab:Preselection}
\begin{tabular}{l r}
\toprule
Requirement                      & Preselection\\
\midrule
Leptons                          & 1\\
Jets                             & $\geq 2$\\
Leading-jet $\pT$ $[\GeV]$       & $> 200$ \\
Leading-jet is central ($|\eta|<2.5$) & Yes\\
Leading-jet is $b$-tagged        & Yes\\
\MET $[\GeV]$                    & $> 120$\\
$|\Delta\phi$(leading jet, \vecMET)$|$ & $\geq 2$\\
\bottomrule
\end{tabular}
\end{table}


\section{Final selection}
\label{sec:finalselection}

After the preselection requirements, events are separated into
signal, control, and validation regions (SRs, CRs, and
VRs) defined by sets of orthogonal event-variable requirements. These
requirements are summarised in Table~\ref{tab:CRs}, with the
variable definitions and motivations given in the following.

\begin{table}%
\centering
\caption{Summary of the selection requirements for the SR compared
with those for the \ttbar and \Wjets CRs and VRs. The
SR, \ttbar CR, and \Wjets CR are split into low-,
middle-, and high-\pTW subregions depending on whether the \pT{}
of the $W$ boson is below 400~\GeV, between 400 and 600~\GeV or
above 600~\GeV, respectively.}
\label{tab:CRs}
\small
\begin{tabular}{l rrrrrr}
\toprule
\diagbox{Requirement}{Region}                      & SR  & \ttbar CR & \ttbar VR & \Wjets CR  & \Wjets VR1 & \Wjets VR2 \\
\midrule
$b$-tagged jets                                    & $\geq 1$ & $\geq 2$ & $\geq 1$ & 1 & 1 & 1\\
Leading-jet $\pT$ $[\GeV]$                           & $> 350$  & $> 200$ & $> 200$ & $> 250$ & $> 250$ & $> 250$ \\
$|\Delta\phi(\text{lepton}, \text{leading jet})|$  & $> 2.5$  & $> 2.5$ & $> 2.5$ & $> 2.5$ & [1.5, 2.5] & [1.5, 2.5] \\
\# additional hard central jets                    & 0        & 0       & $\geq 1$ & -  & - & - \\
\# forward jets ($\pT > 40~\GeV$)                  & $\geq 1$ & 0       & $\geq 1$ & 0   & $\geq 1$ & 0 \\
\bottomrule
\end{tabular}
\end{table}

\subsection{Signal region}

In each signal region event,
the $b$-tagged highest-$\pT$ jet must have $\pt > 350~\GeV$.
To exploit the low multiplicity of high-$\pT$ jets in the signal
process relative to the $\ttbar$ background, events are rejected if they
contain an additional central ($|\eta|<2.5$) jet that has a $\pT > 75~\GeV$ and is
either close to ($\dR < 1.2$) or nearly back-to-back with ($\dR > 2.7$)
the leading jet.

An azimuthal separation of $|\Delta\phi(\text{lepton}, \text{leading jet})| > 2.5$ is required between
the leading jet and the lepton because the leptonically decaying $W$ boson
from a heavy-$Q$ decay recoils against the $b$-quark.

Finally, like in $t$-channel single-top production, the single
production of VLQs gives rise to a forward jet,
so only events with at least one forward jet ($2.5 < |\eta| < 4.5$) with $\pt > 40~\GeV$
are considered.

Figure~\ref{fig:SR_1lep_prefit} shows the reconstructed VLQ
invariant-mass distribution, $\mQreco$, for data events, the expected
SM events (pre-fit), and a VLQ signal in the SR. Events
with \enquote{Fakes} are estimated as detailed in
Section~\ref{sec:multijet}. Weight corrections, obtained as described
in Section~\ref{sec:reweighting}, are applied to the \ttbar and \Wjets
MC samples. \Wjets{} background is separated into light-flavour (\WLF)
and heavy-flavour (\WHF, with jets from $c$,$b$ quarks) components.
For the final fit, the SR is split into three subregions based on the
reconstructed $W$-boson \pT: \enquote{low} with $\pTW <
400~\GeV$, \enquote{middle} with $400~\GeV < \pTW < 600~\GeV$,
and \enquote{high} with $\pTW > 600~\GeV$. This improves the quality
of the results, separating the better modelled, highly populated, low
$\pTW$ region from the other two regions.

\begin{figure}[tb]
\begin{center}
\includegraphics[width=0.42\textwidth]{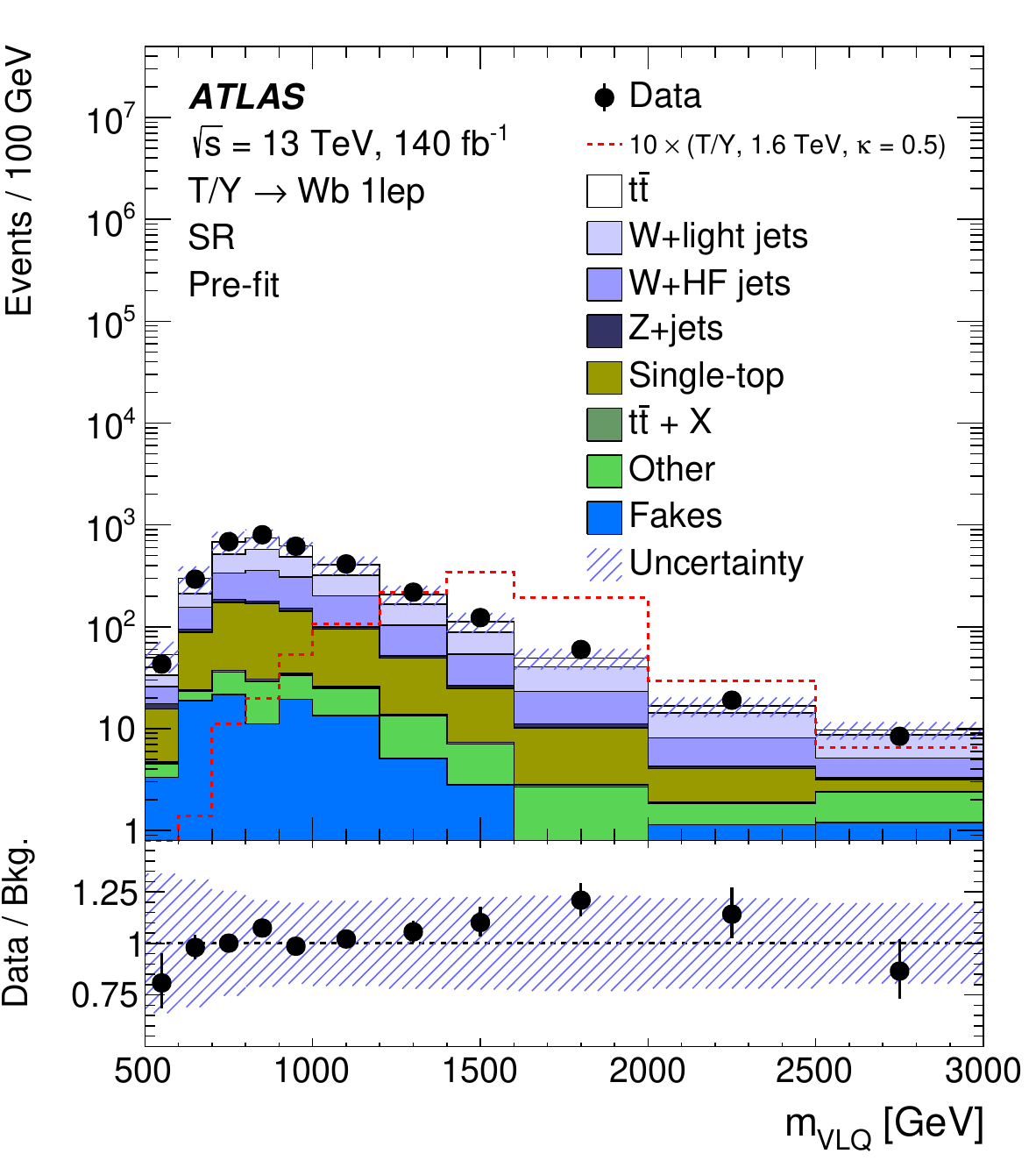}
\caption{Pre-fit VLQ mass distribution in the
signal region. The signal shown here is for a $T/Y$ VLQ with
$\kappa = 0.5$ and a mass of 1600~\GeV, and is presented multiplying
by ten its expected cross-section for clarity.
\enquote{Other} includes diboson and other final states containing a top quark.
The error band
contains the statistical and systematic uncertainties added in
quadrature. Weight corrections are applied to the \ttbar
and \Wjets MC samples (see Section~\ref{sec:reweighting}).
\hfill
\label{fig:SR_1lep_prefit}}
\end{center}
\end{figure}

\subsection{Control regions}

The normalisation of \Wjets and \ttbar processes is partially
constrained by fitting the predicted yields to data in CRs enriched in
either \Wjets or \ttbar events.  As with the SR, each of the two CRs
in Table~\ref{tab:CRs} is split into three subregions
based on $W$-boson \pT: \enquote{low}, \enquote{middle} and
\enquote{high} with $\pTW$ values below 400~\GeV, between
400 and 600~\GeV, and above 600~\GeV respectively.

In addition to the preselection, the selection requirements common
to the three \Wjets CR subregions are that each event must have exactly one
$b$-tagged jet with $\pT > 250~\GeV$, an azimuthal separation of
$|\Delta\phi| > 2.5$ between the lepton and the leading jet, and no
forward jets. These requirements ensure
orthogonality with the SR and increase the fractional \Wjets
contribution. No veto is placed on additional central jets.

In addition to the preselection, the selection requirements common to
the three \ttbar CR subregions are that each event must contain at least two
$b$-tagged jets, with the leading-jet \pT{} greater than 200~\GeV. As
in the SR, there must be no additional hard central jets satisfying
the $\dR < 1.2$ or $\dR > 2.7$ distance criteria, and
the lepton and leading jet must be azimuthally separated by
$|\Delta\phi| > 2.5$. A forward jet veto is applied.

\subsection{Validation regions}

In addition to the CRs, \Wjets and \ttbar VRs are defined in order to
verify that the main backgrounds are well modelled.

Two VRs are defined for \Wjets events: VR1 and VR2. For each, events must
pass the preselection and have exactly one $b$-tagged jet.
Furthermore, the $b$-tagged jet must have a transverse momentum $\pT >
250~\GeV$, but rather than having a very large azimuthal separation
from the lepton
it must satisfy $1.5 < |\Delta\phi| < 2.5$. At least one forward jet
must be present for VR1, whereas forward jets are vetoed for VR2.
VR2 has a larger number of events, and it covers different
kinematic regions.

The \ttbar VR requires at least one $b$-tagged jet, the leading-jet
\pT\ must be greater than 200~\GeV, there must be at least one
additional central high-$\pT$ jet at a distance $\dR < 1.2$ or
$\dR > 2.7$ from the leading jet, and a $|\Delta\phi| > 2.5$
separation is required between the lepton and the leading jet.

\subsection{Estimation of multijet background}
\label{sec:multijet}

Multijet production results in hadrons, photons and non-prompt leptons
that may satisfy the lepton selection criteria and give rise to so-called
\enquote{non-prompt and fake} lepton backgrounds.  The multijet
background's normalisation and shape in the $\mVLQ$ distributions are
estimated with a data-driven method, referred to as the matrix
method.  This approach uses the
efficiencies for leptons selected using \enquote{loose} requirements (\enquote{loose}
leptons) to pass the default \enquote{tight} lepton selection requirements. The
efficiencies are obtained in dedicated control regions enriched in
real leptons or in non-prompt and fake leptons, and applied to events
selected with either the \enquote{loose} or \enquote{tight} lepton definition, to obtain
the fraction of multijet events. The region enriched in real leptons
is composed of events where the invariant mass $m_{\ell\ell}$ reconstructed
from two lepton candidates in the final state, matches the $Z$-boson
invariant mass with $|m_{\ell\ell} - m_Z| < 0.1 \times m_Z$. The default
fake-enriched region for electrons is defined
requiring $\mtw < \SI{20}{\GeV}$ and $\met
+ \mtw < \SI{60}{\GeV}$.\footnote{The quantity $\mtw$ is built as the
invariant mass of the $\ell\nu$ pair, neglecting the $z$-component.} As
a muon fake-enriched region, events with a single muon whose impact
parameter significance satisfies $d_0/\sigma(d_0) > 5$ are
used. Full details can be found in
Refs.~\cite{EXOT-2014-12}.


%
\section{Reweighting of the \Wjets and top-quark backgrounds}
\label{sec:reweighting}


The simulation of the main backgrounds, from \ttbar{} and \Wjets{}
production, is susceptible to mismodelling in the highly energetic
phase-space probed by this analysis, particularly in the modelling
of QCD radiation.
A set of theory and data-driven corrections was hence applied to
the \ttbar{} and \Wjets{} MC samples to improve their accuracy.

Before deriving the data-driven reweightings, the MC samples were
reweighted at generator level to the latest high-precision theory
predictions~\cite{Czakon:2017wor}:
the \Wjets{} background to NLO precision in the electroweak couplings;
the \ttbar background to NNLO QCD precision in the two-dimensional (2D) space of top/anti-top \pT{} and \mtt.

A data-driven reweighting was then used to correct the residual shape
mismodelling in \ttbar{} and \Wjets{} distributions, as described in
the steps listed below.  For the \Wjets{} background, the final fit
provides the overall normalisation, separately for the light-flavour
and heavy-flavour components (see Section~\ref{sec:result}).
\begin{enumerate}
\item A \Wjets enriched region was selected by applying the same
selection cuts as in the preselection, but requiring the number of
$b$-tagged jets to be zero (making it orthogonal to all other
selection regions). This region is depleted of other backgrounds,
with $> 90\%$ of selected events expected to be \Wjets, and hence
mismodelling seen there can be attributed to the \Wjets background.
\item The scaling factors for \Wjets reweighting were determined bin by bin as a
function of both the number of jets and the transverse momentum of
the reconstructed $W$ boson, \pTW.
This \Wjets 2D reweighting function was defined inclusively for both
the \WHF and \WLF components, assuming the same mismodelling,
and fitted to data after subtracting the other (subleading) backgrounds'
expected contributions.
\item This derived reweighting was then applied to the \Wjets{}
MC samples in the preselection region.
\item The preselection region has a \ttbar{}+$Wt$ purity of over 80\%.
Given the similar nature of \ttbar{} and single-top $Wt$ production,
a single \ttbar{}+$Wt$ reweighting function was used. This was
determined bin by bin in the preselection region, as a 2D function
of the leading-jet \pt{} and the leptonic \pTW{}, after subtracting
the expected contributions of other backgrounds, as in the case
of \Wjets reweighting.
\item Finally, the \ttbar{}+$Wt$ reweighting function was applied to
those MC samples in the preselection region.
\end{enumerate}

The weights applied do not deviate markedly from 1 for the majority of
2D bins. For \Wjets reweighting, the scaling factors are between
0.7 and 2.1, while for \ttbar{}+$Wt$ reweighting,
they are between 0.15 and 1.
To show the effect of the combined reweighting,
Figures~\ref{tt_CR2:final_reweighting_mVLQ_comparison} and
\ref{tt_CR2:final_reweighting_leadjetpt_comparison} display the
\ttbar control region $\mQreco$ and leading-jet \pt distributions,
respectively, before and after reweighting.

\begin{figure}[htbp]
\subfigure[]{%
\includegraphics[width=0.48\textwidth]{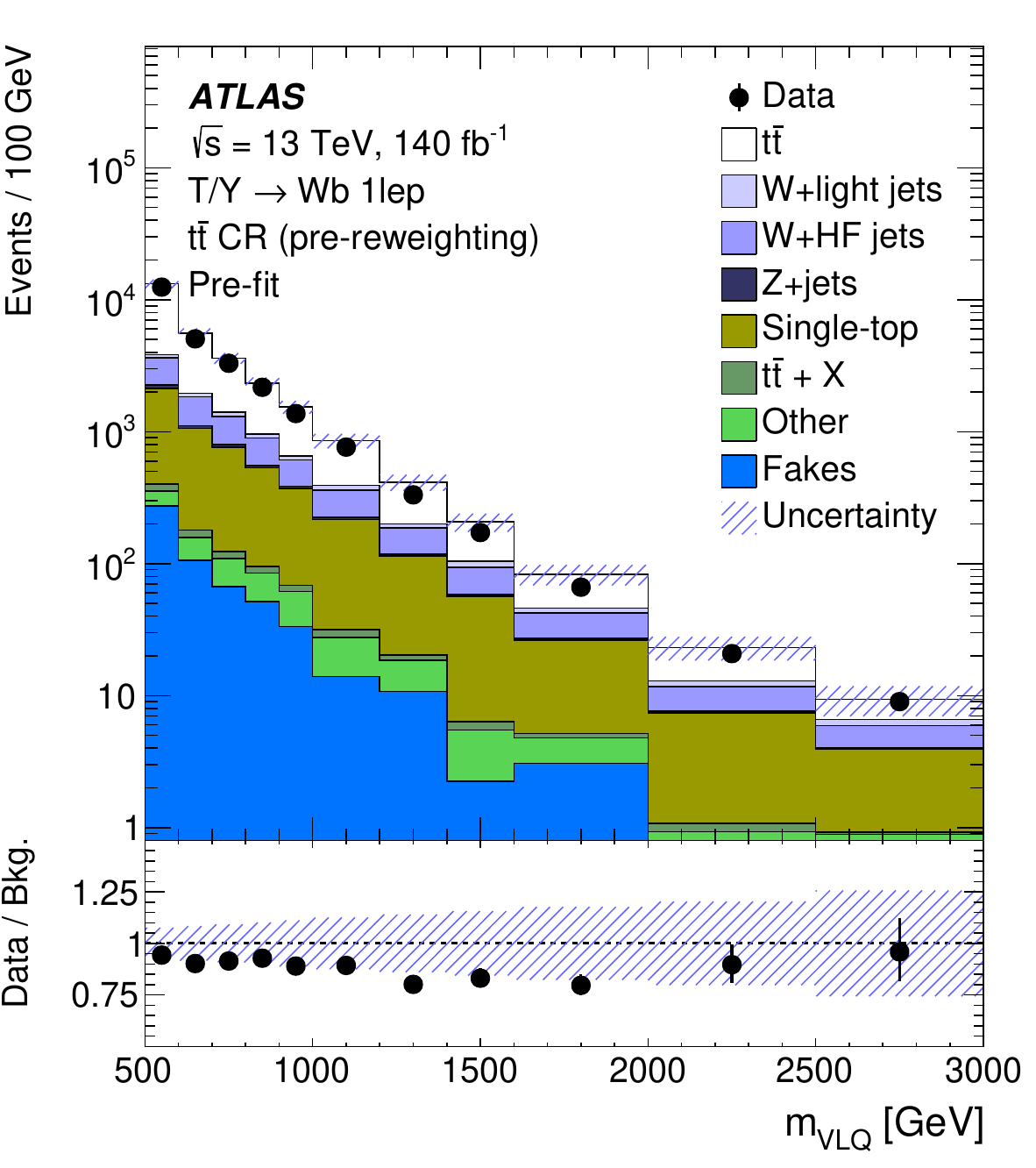}%
\label{tt_CR2:NoRW_reweight_mVLQ_B}%
}
\hfill
\subfigure[]{%
\includegraphics[width=0.48\textwidth]{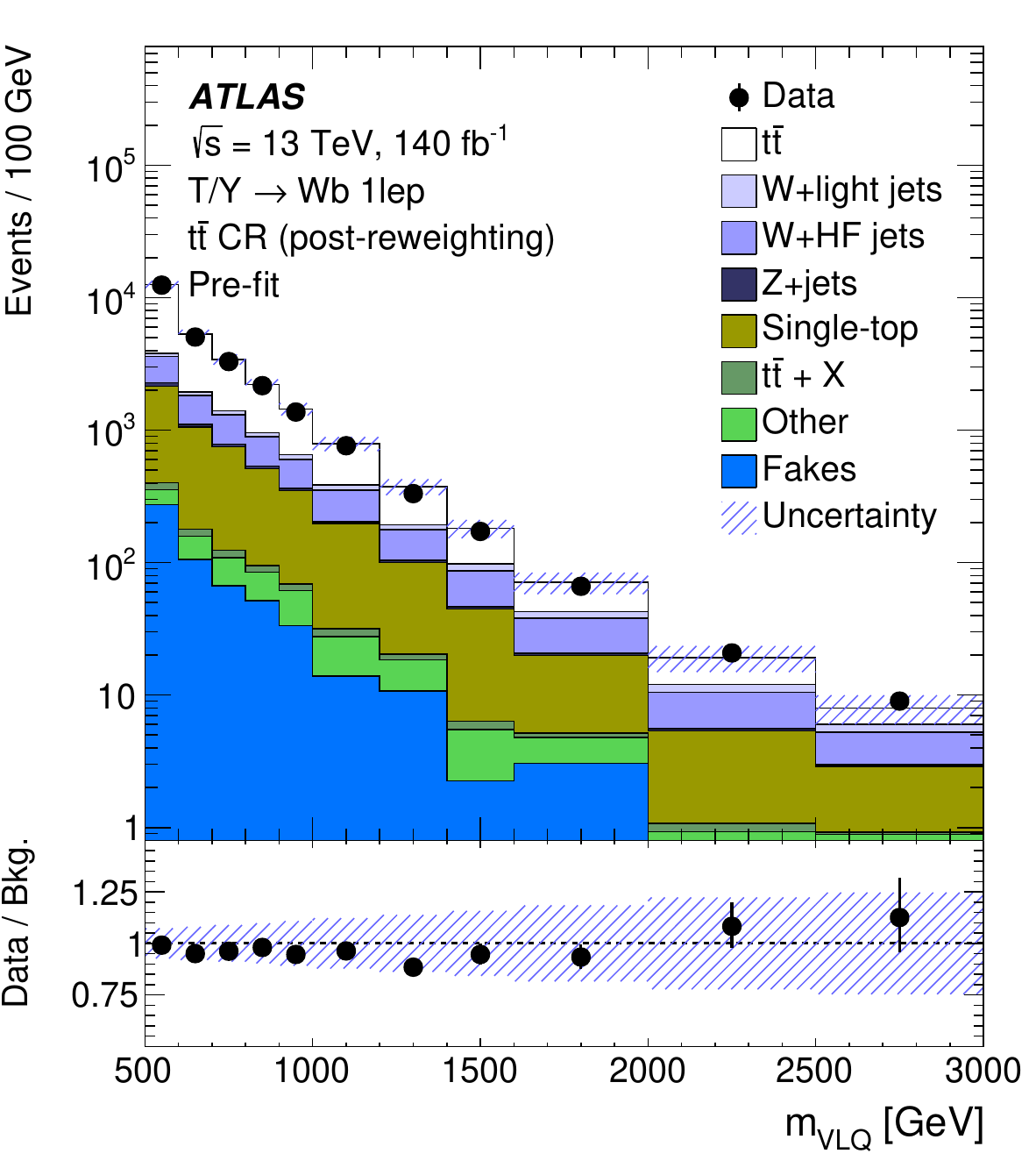}%
\label{tt_CR2:FinalRW_reweight_mVLQ_B}%
}
\caption{$\mQreco$ distribution in the \ttbar CR
\subref{tt_CR2:NoRW_reweight_mVLQ_B} before and
\subref{tt_CR2:FinalRW_reweight_mVLQ_B} after reweighting. Shaded bands include statistical and systematic uncertainties, but not those related to the reweighting procedure. The last bin includes overflow.}
\label{tt_CR2:final_reweighting_mVLQ_comparison}
\end{figure}
\begin{figure}[htbp]
\subfigure[]{%
\includegraphics[width=0.48\textwidth]{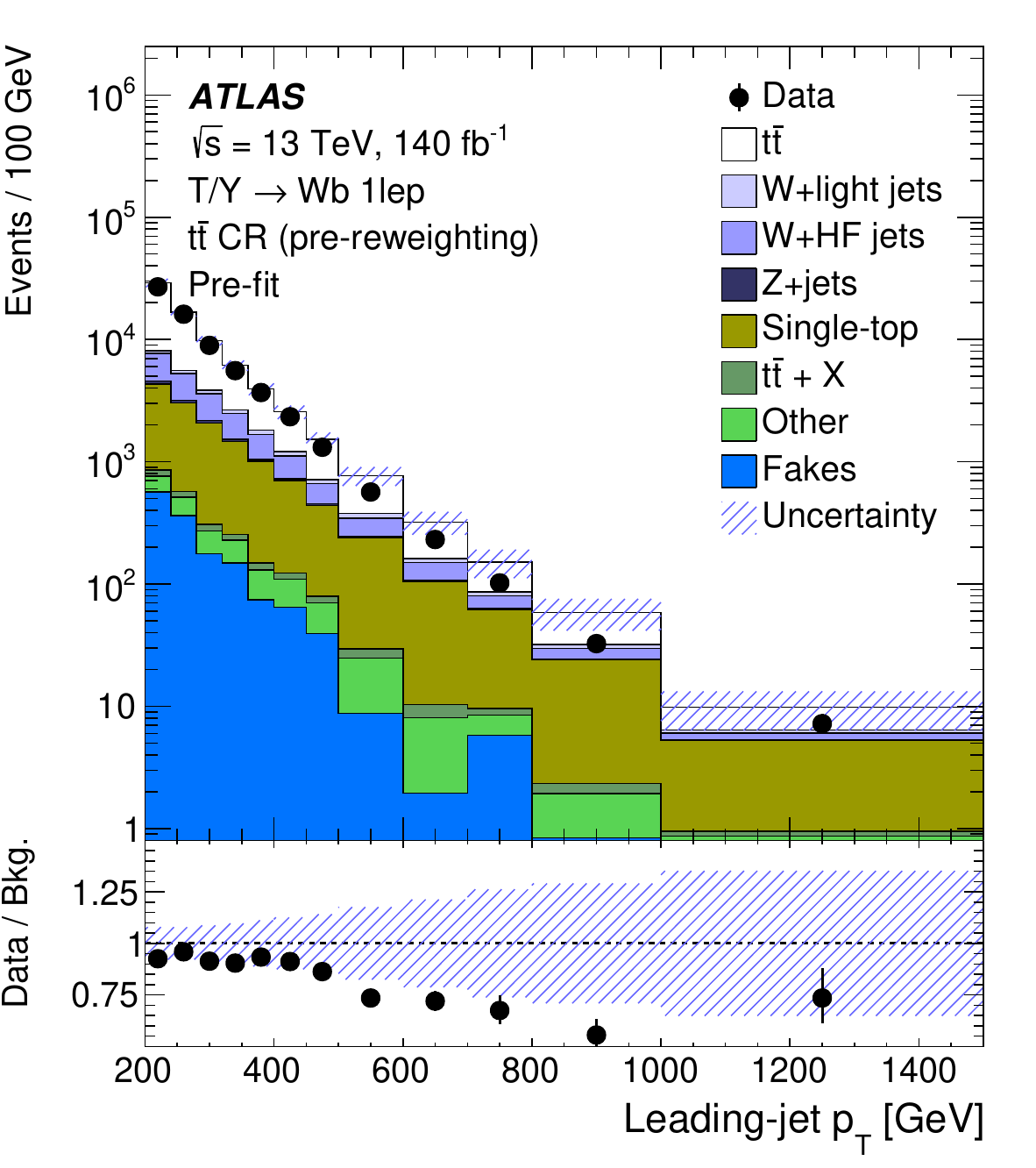}%
\label{tt_CR2:NoRW_reweight_leadjetpt}%
}
\hfill
\subfigure[]{%
\includegraphics[width=0.48\textwidth]{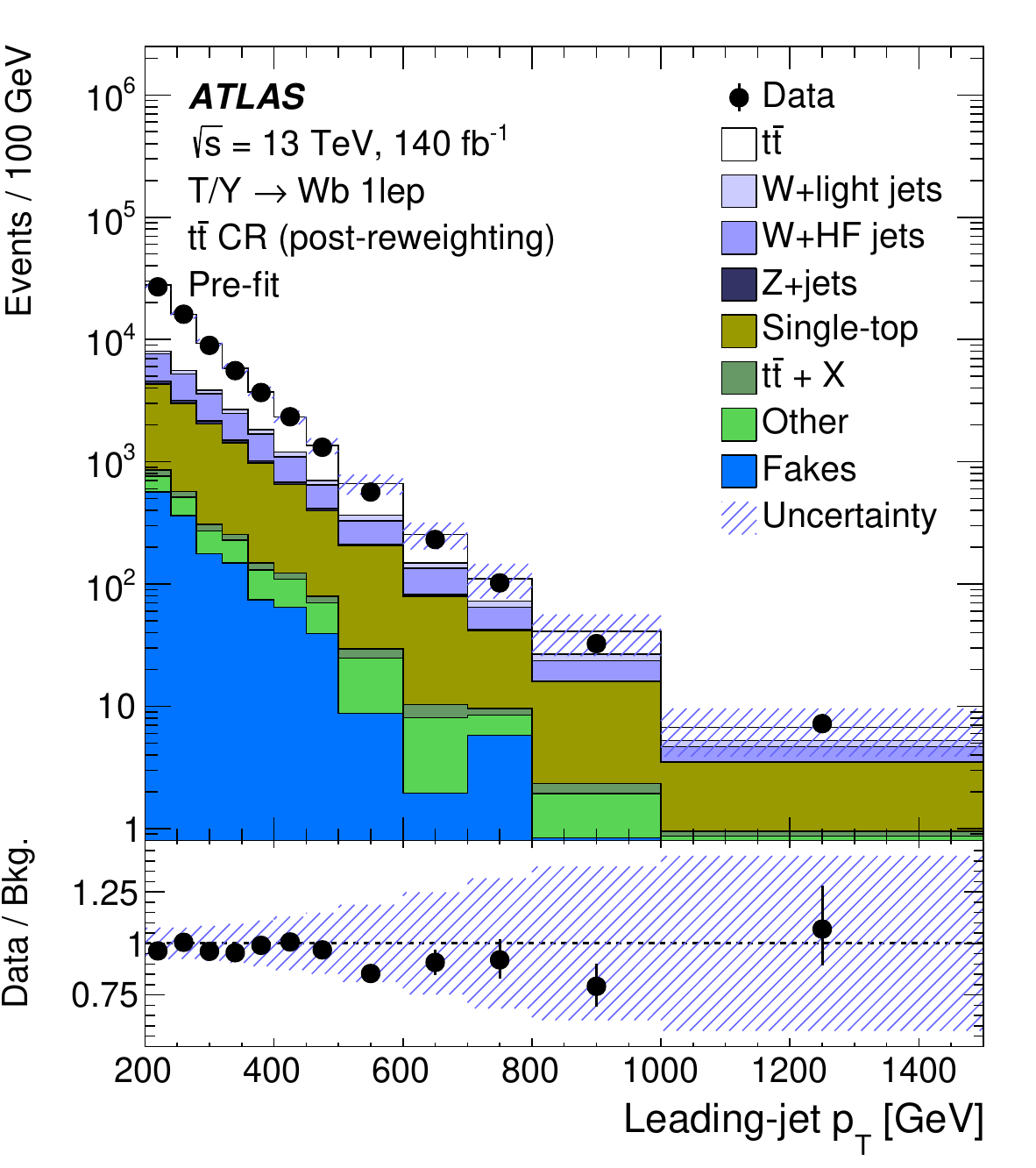}%
\label{tt_CR2:FinalRW_reweight_leadjetpt}%
}
\caption{Leading-jet \pt\ distribution in the \ttbar CR
\subref{tt_CR2:NoRW_reweight_leadjetpt} before and
\subref{tt_CR2:FinalRW_reweight_leadjetpt} after reweighting. Shaded bands include statistical and systematic uncertainties, but not those related to the reweighting procedure. The last bin includes overflow.}
\label{tt_CR2:final_reweighting_leadjetpt_comparison}
\end{figure}

Tables~\ref{tab:wj_CR_yields}--\ref{tab:tt_CR_yields} report the event
yields, with all statistical and systematic uncertainties included, in
the \Wjets and \ttbar control regions, and in the signal region before
the fit. \enquote{Other} includes diboson and other final states containing a
top quark. For a 1.4--2.4~\TeV vector-like $Y$ quark, with a coupling
$\kappa=0.5$, the signal contamination in the high-\pTW, \Wjets
and \ttbar CRs ranges from 0.1\% to 3\% and from 2.5\% to 19\%,
respectively.

\begin{table}[htbp]
\begin{center}
\caption{\Wjets control region yields after reweighting the \Wjets and top backgrounds. Expected $Y$ signals, for coupling $\kappa=0.5$ and three masses, are also listed.}
\label{tab:wj_CR_yields}
\begin{tabular}{l
S[table-format=5.2(2)]
S[table-format=5.2(2)]
S[table-format=4.3(2)]}
\toprule
& {\Wjets low-\pTW} & {\Wjets mid-\pTW} & {\Wjets high-\pTW} \\
\midrule{\ttbar}            & 7110 \pm 890 & 1930 \pm 310 & 210 \pm 100 \\
{\WLF jets}         & 12200 \pm 1500 & 4140 \pm 620 & 1070 \pm 270 \\
{\WHF jets}         & 12510 \pm 850 & 3020 \pm 380 & 610 \pm 140 \\
{$Z$+jets}          & 854 \pm 80 & 129 \pm 55 & 22.2 \pm 4.5 \\
{Single top}        & 2000 \pm 790 & 700 \pm 340 & 140 \pm 130 \\
{$\ttbar V$}        & 33.2 \pm 4.3 & 16.7 \pm 2.8 & 5.1 \pm 1.0 \\
{$\ttbar H$}        & 7.10 \pm 0.94 & 2.04 \pm 0.33 & 0.42 \pm 0.11 \\
{Other}           & 954 \pm 74 & 400 \pm 51 & 130 \pm 22 \\
{Fakes ($e$)}       & 510 \pm 250 & 30 \pm 15 & 29 \pm 14 \\
{Fakes ($\mu$)}     & 940 \pm 470 & 91 \pm 45 & 12.2 \pm 6.1 \\
\midrule{$Y$ (1.4~\TeV, $\kappa = 0.5$)} & 228 \pm 10 & 210.2 \pm 8.8 & 65.7 \pm 2.8 \\
{$Y$ (1.9~\TeV, $\kappa = 0.5$)} & 97.8 \pm 4.4 & 31.8 \pm 1.5 & 9.50 \pm 0.36 \\
{$Y$ (2.4~\TeV, $\kappa = 0.5$)} & 31.8 \pm 1.5 & 6.89 \pm 0.29 & 2.073 \pm 0.082 \\
\midrule{Total background}  & 37100 \pm 2500 & 10460 \pm 990 & 2230 \pm 380 \\
\midrule{Data}              & 38191 & 11046 & 2425 \\
\bottomrule
\end{tabular}
\end{center}
\end{table}

\begin{table}[htbp]
\begin{center}
\caption{\ttbar control region yields after reweighting the \Wjets and top backgrounds. Expected $Y$ signals, for coupling $\kappa=0.5$ and three masses, are also listed.}
\label{tab:tt_CR_yields}
\begin{tabular}{l
S[table-format=5.3(2)]
S[table-format=5.2(2)]
S[table-format=4.2(2)]}
\toprule
& {\ttbar CR low-\pTW} & {\ttbar CR mid-\pTW} & {\ttbar CR high-\pTW} \\
\midrule{\ttbar}            & 16500 \pm 1000 & 1040 \pm 150 & 119 \pm 57 \\
{\WLF jets}         & 530 \pm 160 & 131 \pm 44 & 35 \pm 14 \\
{\WHF jets}         & 3140 \pm 370 & 559 \pm 87 & 113 \pm 31 \\
{$Z$+jets}          & 264 \pm 42 & 20.7 \pm 4.5 & 2.4 \pm 1.2 \\
{Single top}        & 3800 \pm 1200 & 642 \pm 430 & 150 \pm 150 \\
{$\ttbar V$}        & 66.5 \pm 7.4 & 11.9 \pm 2.2 & 2.93 \pm 0.73 \\
{$\ttbar H$}        & 29.7 \pm 3.2 & 3.22 \pm 0.48 & 0.52 \pm 0.11 \\
{Other}             & 224 \pm 28 & 59 \pm 17 & 13.0 \pm 5.8 \\
{Fakes ($e$)}       & 220 \pm 110 & 5.3 \pm 2.7 & 2.4 \pm 1.2 \\
{Fakes ($\mu$)}     & 370 \pm 180 & 20 \pm 10 & 2.1 \pm 1.1 \\
\midrule{$Y$ (1.4~\TeV, $\kappa = 0.5$)} & 25.7 \pm 1.2 & 71.0 \pm 3.4 & 82.2 \pm 4.8 \\
{$Y$ (1.9~\TeV, $\kappa = 0.5$)} & 3.35 \pm 0.20 & 11.68 \pm 0.65 & 33.7 \pm 1.8 \\
{$Y$ (2.4~\TeV, $\kappa = 0.5$)} & 0.703 \pm 0.051 & 2.12 \pm 0.11 & 11.49 \pm 0.69 \\
\midrule{Total background}  & 25200 \pm 1800 & 2500 \pm 500 & 440 \pm 170 \\
\midrule{Data}              & 24582 & 2348 & 363 \\
\bottomrule
\end{tabular}
\end{center}
\end{table}


%

%
\section{Systematic uncertainties}
\label{sec:systematics}

Several sources of systematic uncertainty in this analysis can affect
the normalisation of the signal and background and/or their
corresponding $\mQreco$ distributions used for the statistical study.
These include the modelling of the detector response,
object-reconstruction algorithms, uncertainties in the theoretical
modelling of the signals and backgrounds, and uncertainties arising
from the limited size of the simulated event samples.

These effects were estimated using various methods described in this
section, and were included in the statistical analysis via nuisance
parameters. The largest systematic uncertainties in the SM
background estimates arise from the jet-energy scale calibration, the
flavour-tagging efficiencies ($b$, $c$ and light), and the background
modelling.

\subsection{Experimental uncertainties}

The estimated uncertainty in the integrated luminosity of the combined 2015--2018 dataset is 0.83\%, obtained via a methodology similar to that
detailed in Ref.~\cite{DAPR-2021-01}.

The efficiency and mistag rates of the flavour-tagging algorithm were
measured using data control samples. Correction factors were then
calculated to align the tagging rates in the simulation with the
efficiencies observed in the data control
samples~\cite{FTAG-2018-01,FTAG-2020-08,FTAG-2019-02}.  These
correction factors were determined only for $b$- and $c$-jets with
$\pT < 300~\GeV$ and light-flavour jets with $\pt < 750~\GeV$.

The uncertainties associated with the measurements used to derive
these correction factors were
factorised into groups of statistically independent sources, including six
independent sources affecting $b$-jets and four independent sources
affecting $c$-jets. Each of these uncertainties has a different
dependence on the jet \pt. Seventeen sources of uncertainty affecting
light-flavour jets are also considered, and these depend on jet \pt and
$\eta$.  An additional relative uncertainty, fully
correlated among these three jet flavours, is applied for extrapolation of
these corrections to jets beyond the \pT-reach of the data-calibration
samples.
An uncertainty related to the application of $c$-jet scale factors to
$\tau$-jets was considered, but was found to have a negligible impact
in this analysis~\cite{PERF-2016-05}.

Additional detector-related systematic uncertainties arise from the
reconstruction and measurement of jets~\cite{JETM-2018-05},
leptons~\cite{TRIG-2018-05,TRIG-2018-01,EGAM-2018-01,MUON-2018-03} and
$\met$~\cite{JETM-2020-03}:
\begin{itemize}
\item Uncertainties associated with jets arise primarily from the
jet-energy scale, jet-energy resolution, and the efficiency of the JVT
requirement. The largest contribution is from the jet-energy scale:
the dependence of this on jet \pt and $\eta$, jet flavour, and
pile-up is encoded in 21 uncorrelated components that are treated
independently in the statistical analysis~\cite{JETM-2018-05}.
\item Uncertainties associated with leptons arise from the trigger,
reconstruction, identification, and isolation efficiencies, as well
as the lepton-momentum scale and resolution. These were studied
using $Z \rightarrow \ell^{+} \ell^{-}$ and $J/\psi \rightarrow
\ell^{+} \ell^{-}$ decays in data.
\item The systematic uncertainty in the \MET reconstruction is
dominated by the uncertainties in the energy calibration and
resolution of reconstructed jets and leptons. They are propagated
to the $\met$ via their inclusion among the uncertainties associated with
those objects~\cite{JETM-2020-03}. Subleading contributions are also included
via the \pt{} scale and resolution uncertainties of reconstructed
tracks that are associated with the hard-scatter vertex but not
matched to any reconstructed objects.
\end{itemize}

The flavour-tagging systematic uncertainties are the leading source
of experimental uncertainty: added in quadrature, they account for
an 8.7\% uncertainty in the expected background yield in the SR.
Other significant detector-specific uncertainties arise from jet-energy
scale uncertainties (a 6.4\% effect on the expected background yield)
and jet-energy resolution uncertainties (a 2.7\% effect on the expected
background yield). The total systematic uncertainty associated with
\MET{} reconstruction is about 0.3\% of the expected SR background
yield.

For the data-driven multijet background, which contributes very little
to the SR and CRs, a 50\% normalisation uncertainty is applied in
order to fully cover discrepancies between the observed data and
the SM expectation in control regions enhanced in multijet background.

\subsection{Theoretical modelling uncertainties}

A number of systematic uncertainties arise in the theoretical
modelling of the signal and background processes detailed in
Section~\ref{sec:samples}, and these were estimated using various strategies.
All samples include theoretical-uncertainty estimates from variations
of the matrix-element factorisation and renormalisation scales
(\muF and \muR), and from standard PDF-set variations that
capture the statistical and modelling uncertainties in the PDF fits.

\subsubsection*{Signal-modelling systematic uncertainties}

The systematic uncertainties in the modelling of the high-mass $Y$/$T$-quark
signal sample which correspond to the choice of PDF set were evaluated
following the PDF4LHC15 prescription~\cite{Butterworth:2015oua}.  No
further systematic uncertainties in the signal modelling, and no
uncertainties in the NLO signal-production cross-section, are
considered.  In addition, a uniform systematic uncertainty of about 2.5\% is
applied to cover small differences in the reconstructed VLQ mass
between signal samples passed through the full simulation of the
detector and signal samples produced with the faster simulation.

\subsubsection*{Background-modelling systematic uncertainties}
\label{sec:bkgd-syst}
For the \ttbar process, NLO+LO QCD effects on the top-quark kinematics
are considered a source of systematic uncertainty.  Three effects were
combined into one nuisance parameterisation to account for systematic
uncertainties in initial-state radiation (ISR): the \POWHEG \hdamp
parameter was varied from the nominal $1.5 m_t$ to $3 m_t$; the scales
\muF and \muR were raised and lowered by factors of two; and the
Var3c set of  parameter variations were applied in the parton shower's
A14 tune~\cite{ATL-PHYS-PUB-2017-007}.
In addition, the QCD final-state radiation (FSR) uncertainty was estimated by
varying the strong coupling constant \alphas in the \Pythia parton shower.

For single top-quark production, the \enquote{diagram subtraction}
scheme~\cite{Frixione:2008yi} was considered as an alternative
to the nominal \enquote{diagram removal} scheme for treating
overlap between \ttbar and $Wt$ final states, with the full
difference between the two methods applied as a shape and
normalisation uncertainty~\cite{White:2009yt}.  The effect of a possible
overestimation of ISR was evaluated by comparing the nominal sample
with one obtained by doubling the renormalisation and factorisation
scales and choosing the \enquote{Var3cDown} weight of the A14
tune~\cite{ATL-PHYS-PUB-2020-023}.  The effect of a possible
underestimation of ISR was evaluated by comparing the nominal sample
with one produced with \hdamp{} increased to 3$m_t$, the renormalisation
and factorisation scales halved, and using the \enquote{Var3cUp}
weight of the A14 tune.  The size of the uncertainty in the modelling
of FSR was estimated by doubling or halving the
renormalisation scale for emissions from the parton shower.

The normalisations of the $\ttbar$ and \Wjets backgrounds are treated
as free parameters in the fit and hence no \emph{a priori} uncertainty is assigned
to them.  The normalisation of the single-top background is assigned
an uncertainty of 6.8\%~\cite{Kidonakis:2010ux}.
For the \Wjets and \Zjets processes modelled using \Sherpa,
systematic uncertainties from the choice of generator were estimated via
comparison with the \MGNLO alternative samples.
The modelling of the \Zjets  and diboson background
processes includes a 5\% effect from their
normalisation to the theoretical NNLO or NLO
cross-section, respectively~\cite{Melnikov:2006kv, Campbell:1999ah,
Anastasiou:2003ds}.  Since both of these backgrounds are very small,
this effect is applied as an uncertainty in the sum of the predicted
\Zjets and diboson backgrounds.

All normalisation uncertainties in the different background processes
are treated as uncorrelated. For background estimates based on
simulations, the largest  theoretical modelling uncertainties are due
to the choice of parton shower and hadronisation model (2\%--4\%),
the choice of generator (about 1\%--3\% in the expected
background yield), and varying the parameters controlling the initial-
and final-state radiation (about 0.1\% in the expected background
yield), where the theoretical modelling uncertainties from $\ttbar$
contribute the most.

The uncertainty from the data-driven reweighting of the \Wjets and
\ttbar simulations, described in Section~\ref{sec:reweighting},
was estimated as 100\% of the difference between those samples with
and without the reweighting factors applied. The reweighting and other
modelling uncertainties are not treated as being correlated between
\WLF and \WHF, so the fit has the freedom to further correct \WLF and
\WHF differently if needed. These uncertainties were separated into independent sources of
systematic uncertainty in the three \pTW subregions for the
statistical fit.


%

%
\section{Results}
\label{sec:result}


\subsection{Statistical interpretation}
\label{sec:StatMethods}

Testing for the presence of a VLQ signal is performed through binned
profile-likelihood fits of the models to the data, incorporating the
previously estimated systematic uncertainties via nuisance
parameters. The inputs to the fit are the binned distributions of
reconstructed VLQ candidate mass $\mVLQ$ in the SR and the two CRs,
each subdivided into low, middle, and high $W$-boson \pt.  An
independent parameter fit was performed for each signal hypothesis --
a combination of VLQ flavour, multiplet, mass and coupling.

The binned likelihood function $\mathcal{L}(\mu,\theta)$ is constructed
as a product of Poisson probability terms over all  $\mVLQ$ bins in the
SR and CRs. For each model point, this function depends on the
signal-strength parameter $\mu$, which multiplies the nominal
theoretical prediction of the signal-production cross-section, and the
set of nuisance parameters $\theta$, which encode the effects of
systematic uncertainties in the signal and background expectations.
The systematic uncertainties are implemented as unit Gaussian
constraints on the nuisance parameters, which interpolate model
variations of the $\mVLQ$ spectrum away from the nominal
estimates, as well as unconstrained scale factors for the nominal
\ttbar and \Wjets background normalisations. Uncertainties in each
bin of the $\mVLQ$ distributions due to the finite size of the
simulation samples are included using dedicated fit parameters
and are propagated to the signal-strength parameter $\mu$.

The nuisance parameters $\theta$ allow variations of the expectations
for signal and background according to the corresponding systematic
uncertainties, and their fitted values $\hat{\theta}$ correspond to
the deviations from the nominal expectations which globally provide
the best fit to the data. This procedure reduces the impact of
systematic uncertainties on the search sensitivity by taking advantage
of the well-populated background-dominated CRs included in the
likelihood fit to provide data-driven constraints from phase-space
regions close to the SR, and allowing the CRs to improve the
description of the data by using non-BSM aspects of the statistical model.

For models where interference plays a role, $\sqrt{\mu}$ replaces
$\mu$ as the signal-strength parameter varied in the fit. This applies to
the signal $S$, interference $I$, and background $B$ as
\begin{equation*}
\mu \cdot S + \sqrt{\mu} \cdot I + B \, ,
\end{equation*}
where the interference term $I$ is obtained as described in
Section~\ref{sec:signal-samples}. Due to the size of the interference
effect, a uniform uncertainty of 2.5\% is assigned to the interference
term.

The test statistic $q_\mu$ is defined as the profile log-likelihood
ratio, $q_\mu =
-2\ln[\mathcal{L}(\mu,\hat{\hat{\theta}}_\mu)/\mathcal{L}(\hat{\mu},\hat{\theta})]$,
where $\hat{\mu}$ and $\hat{\theta}$ are the values of the parameters
that maximise the likelihood function (with the constraint $0\leq
\hat{\mu} \leq \mu$), and $\hat{\hat{\theta}}_\mu$ are the values of
the nuisance parameters that maximise the likelihood function
conditional on a fixed value of $\mu$. For models including
interference, the
nominal-$\mu$ test statistic $q_{1,0} =
-2\ln[\mathcal{L}(1,\hat{\hat{\theta}}_1)/\mathcal{L}(0,
\hat{\hat{\theta}}_0)]$ (with doubly conditional $\theta$
optimisation), is used instead to stabilise the profiling~\cite{EXOT-2020-25}.

In the absence of any significant deviation from the background
expectation, $q_\mu$ and $q_0$ are used in the asymptotic
approximation of the $\mathrm{CL}_\text{s}$
method~\cite{Junk:1999kv,Read:2002hq,Cowan:2010js} to set an upper
limit on the signal-production cross-section times branching
ratio. For a given signal scenario, values of the production
cross-section (parameterised by $\mu$) which yield
$\mathrm{CL}_\text{s} < 0.05$ are excluded at 95\% confidence level
(CL). In interference-sensitive model interpretations using the
$q_{1,0}$ test statistic, the limit bands are calculated using the
modified method described in Ref.~\cite{EXOT-2020-25}.

\subsection{Background-only fit}

\begin{figure}[tbhp]
\centering
\subfigure[]{
\includegraphics[width=0.32\textwidth]{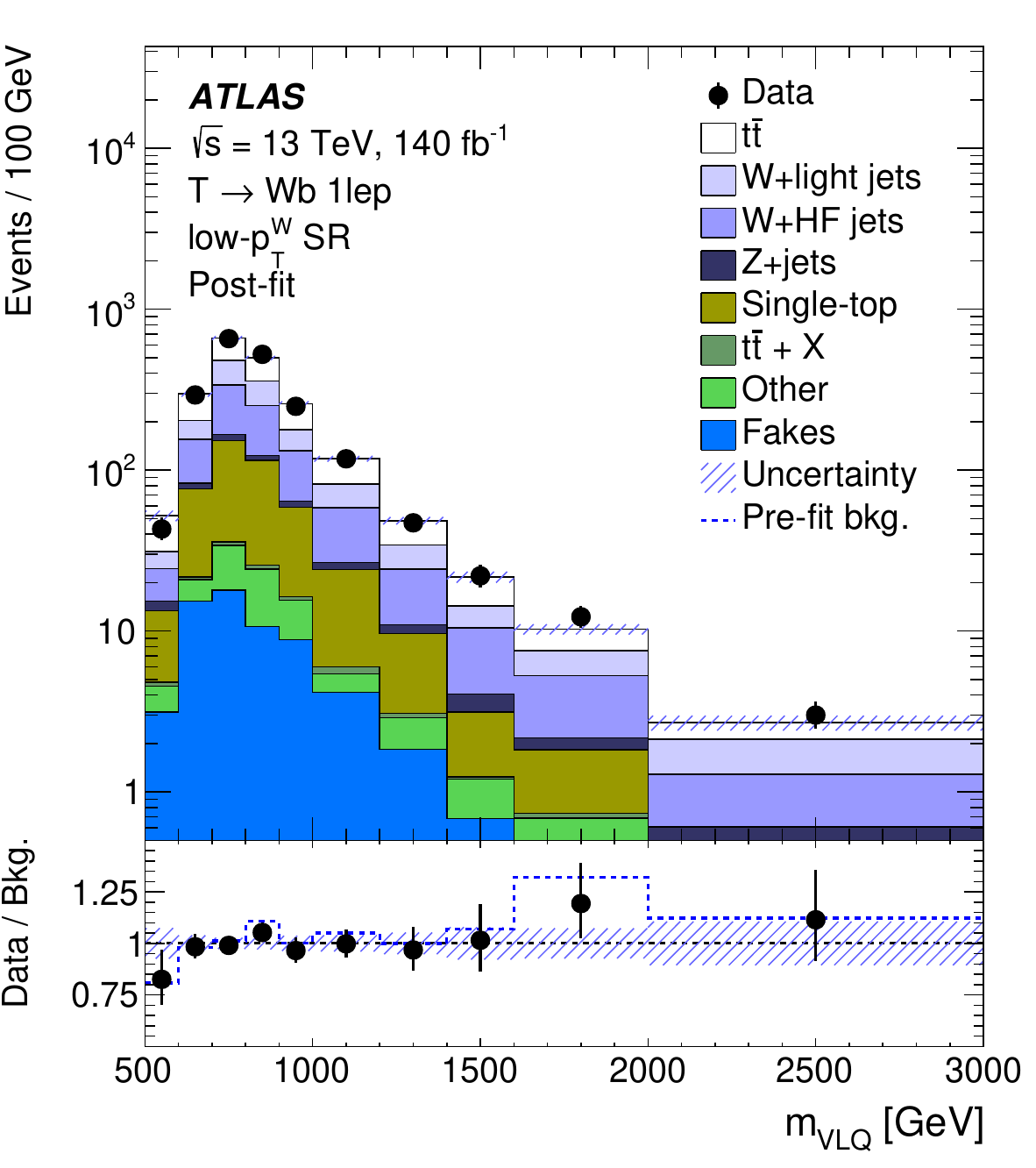}%
\label{Fig:sr_lowwpt_postfit_unblinded}%
}%
\hfill
\subfigure[]{
\includegraphics[width=0.32\textwidth]{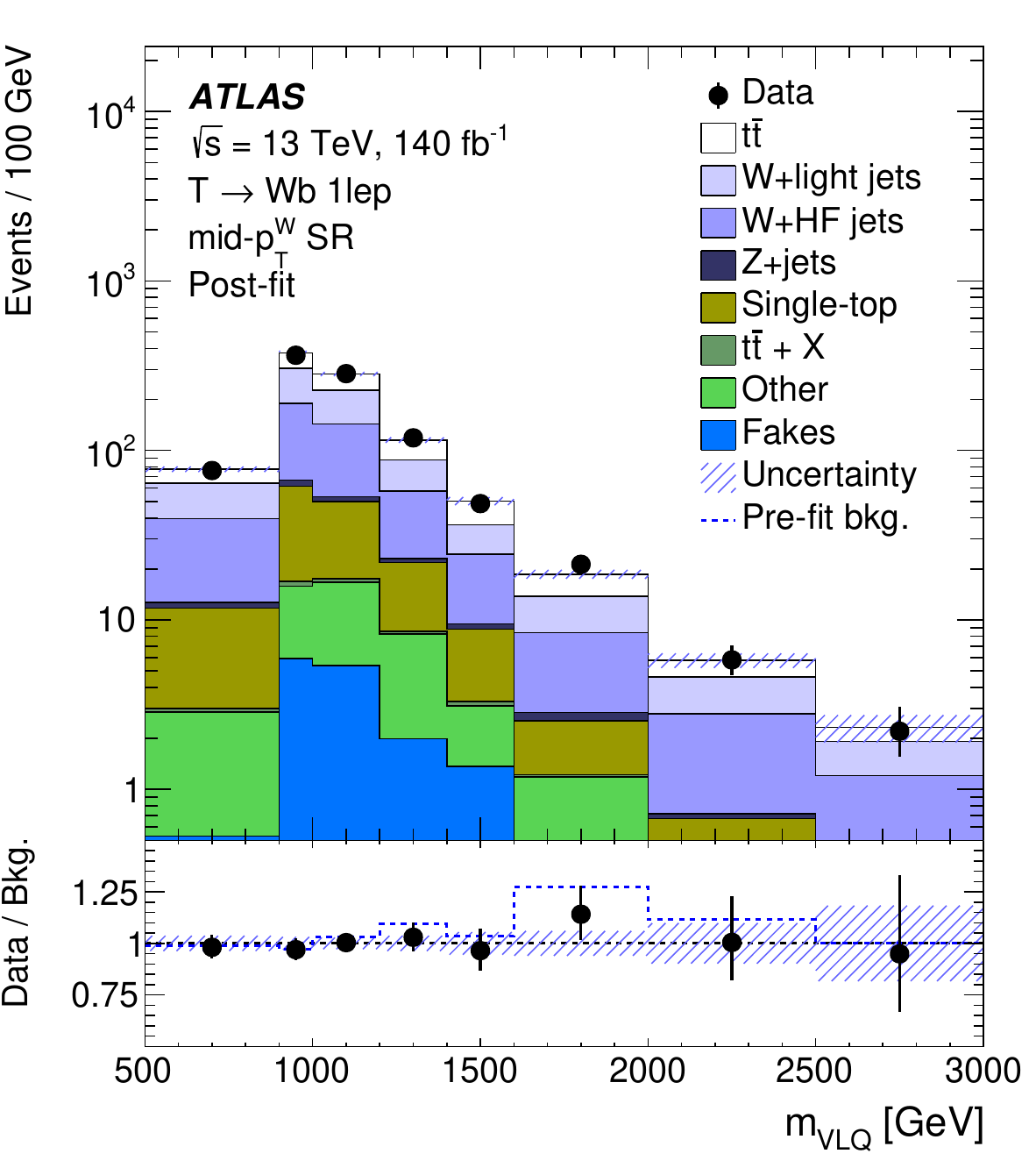}%
\label{Fig:sr_midwpt_postfit_unblinded}%
}%
\hfill
\subfigure[]{
\includegraphics[width=0.32\textwidth]{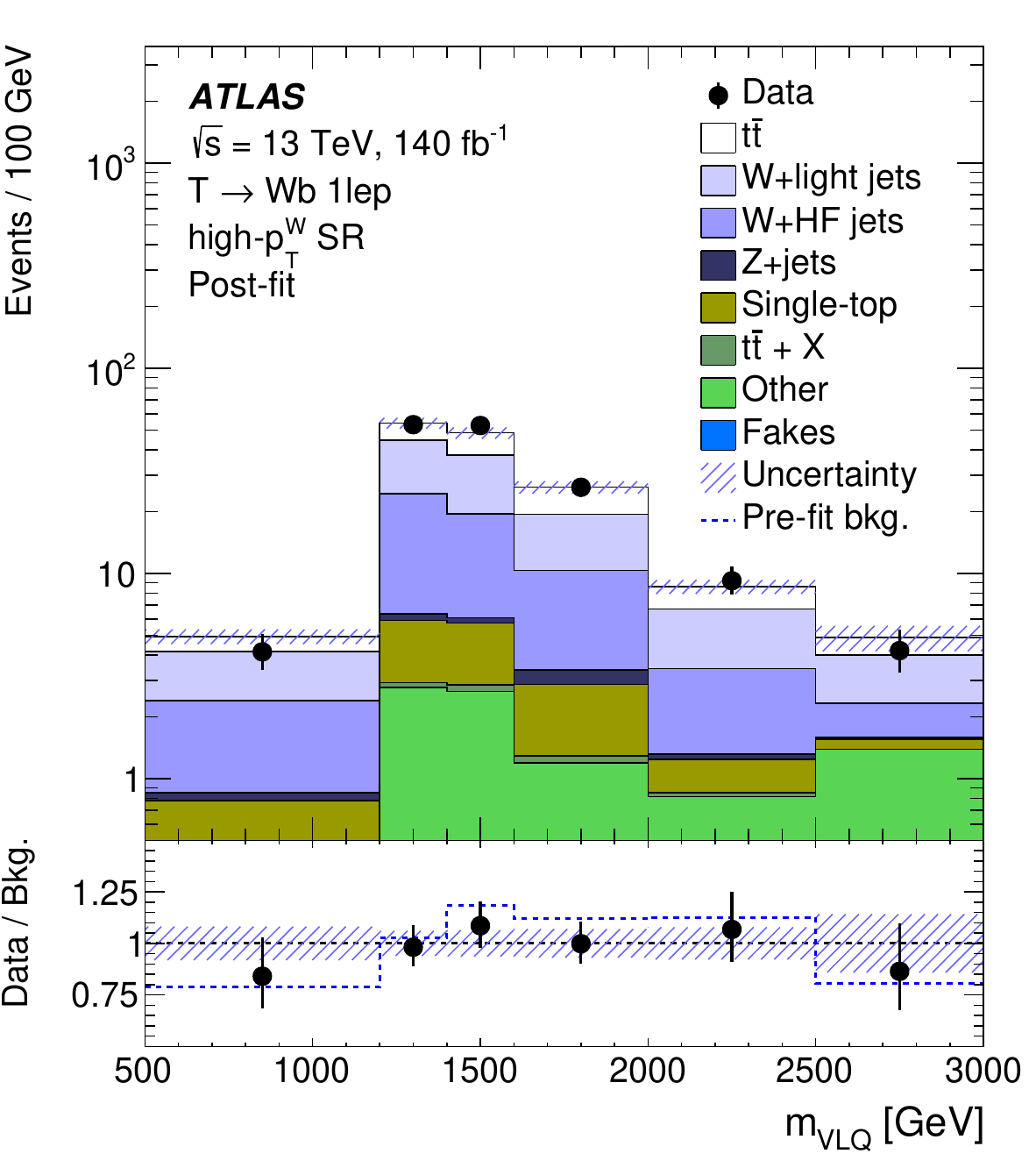}%
\label{Fig:sr_highwpt_postfit_unblinded}%
}\\
\subfigure[]{
\includegraphics[width=0.32\textwidth]{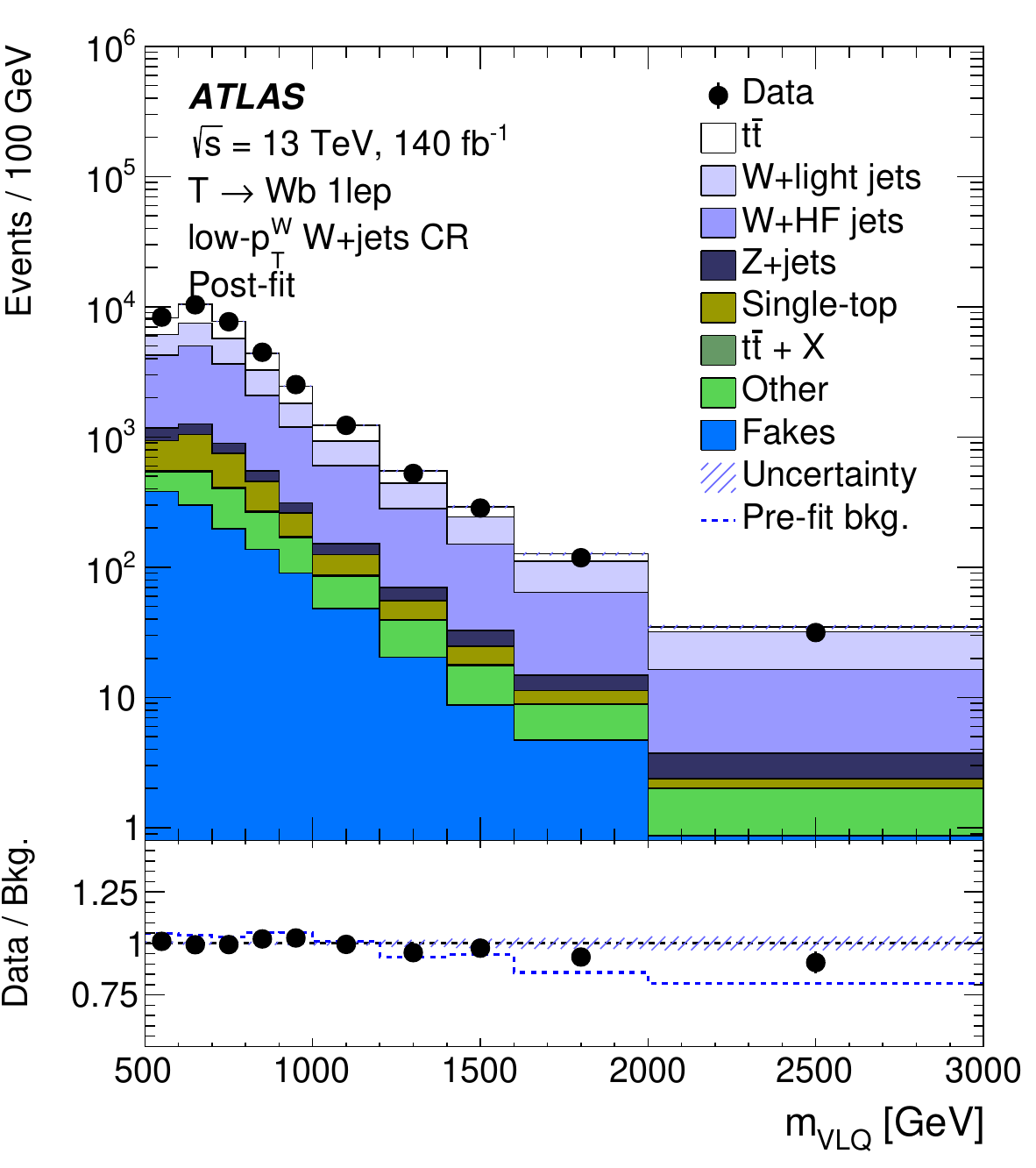}%
\label{Fig:wjcr_lowwpt_postfit_unblinded}%
}%
\hfill
\subfigure[]{
\includegraphics[width=0.32\textwidth]{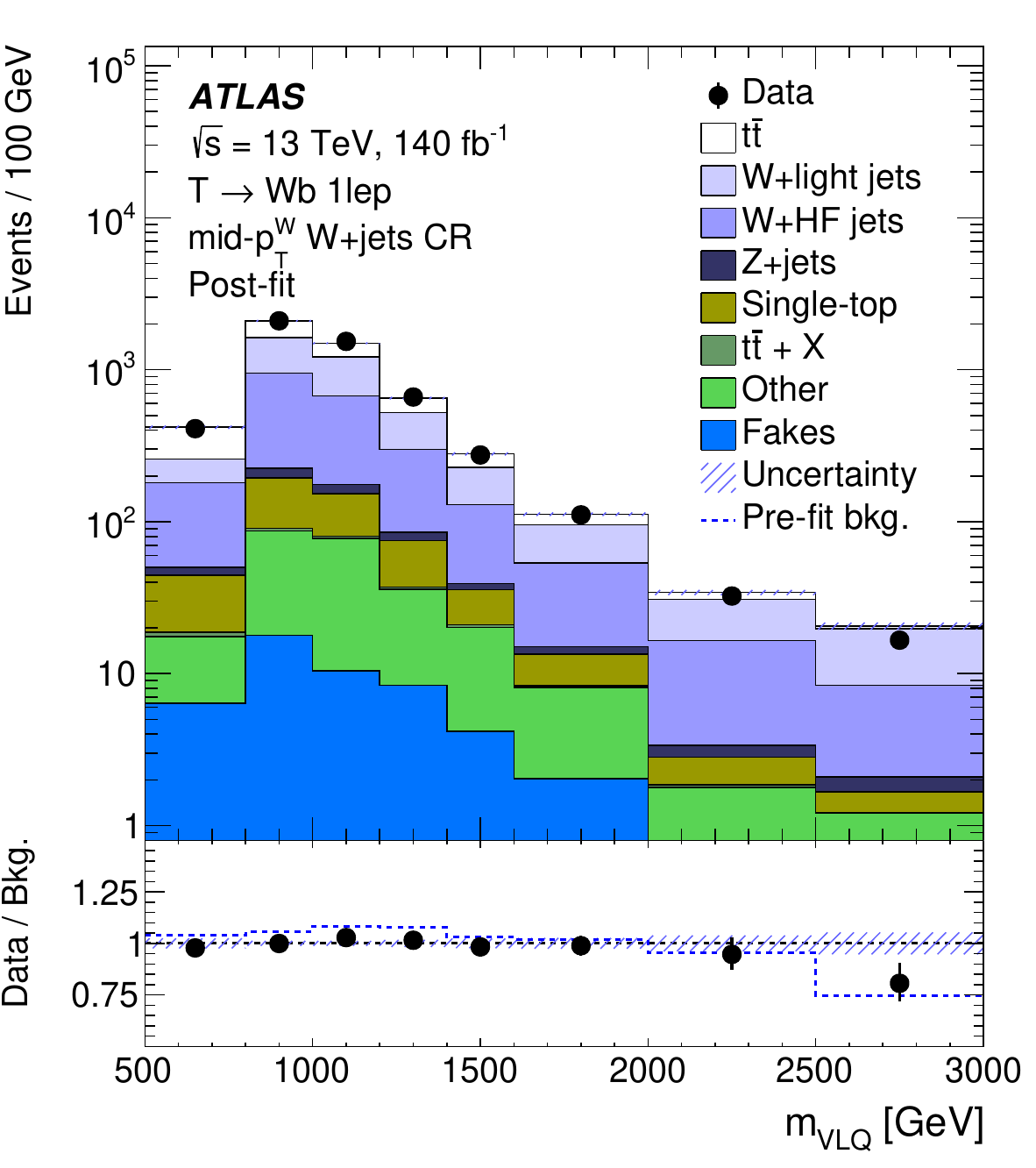}%
\label{Fig:wjcr_midwpt_postfit_unblinded}%
}%
\hfill
\subfigure[]{
\includegraphics[width=0.32\textwidth]{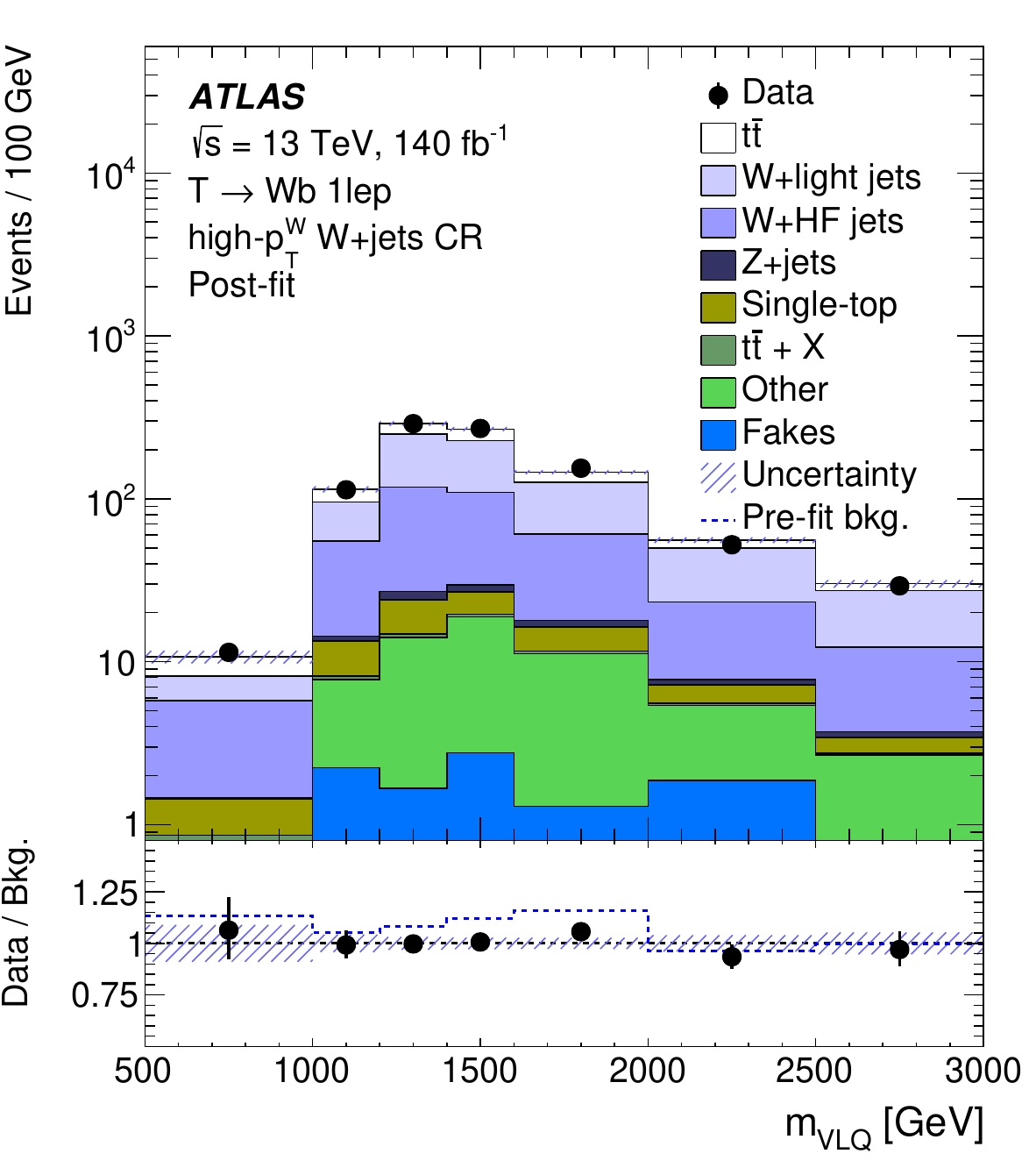}%
\label{Fig:wjcr_highwpt_postfit_unblinded}%
}\\
\subfigure[]{
\includegraphics[width=0.32\textwidth]{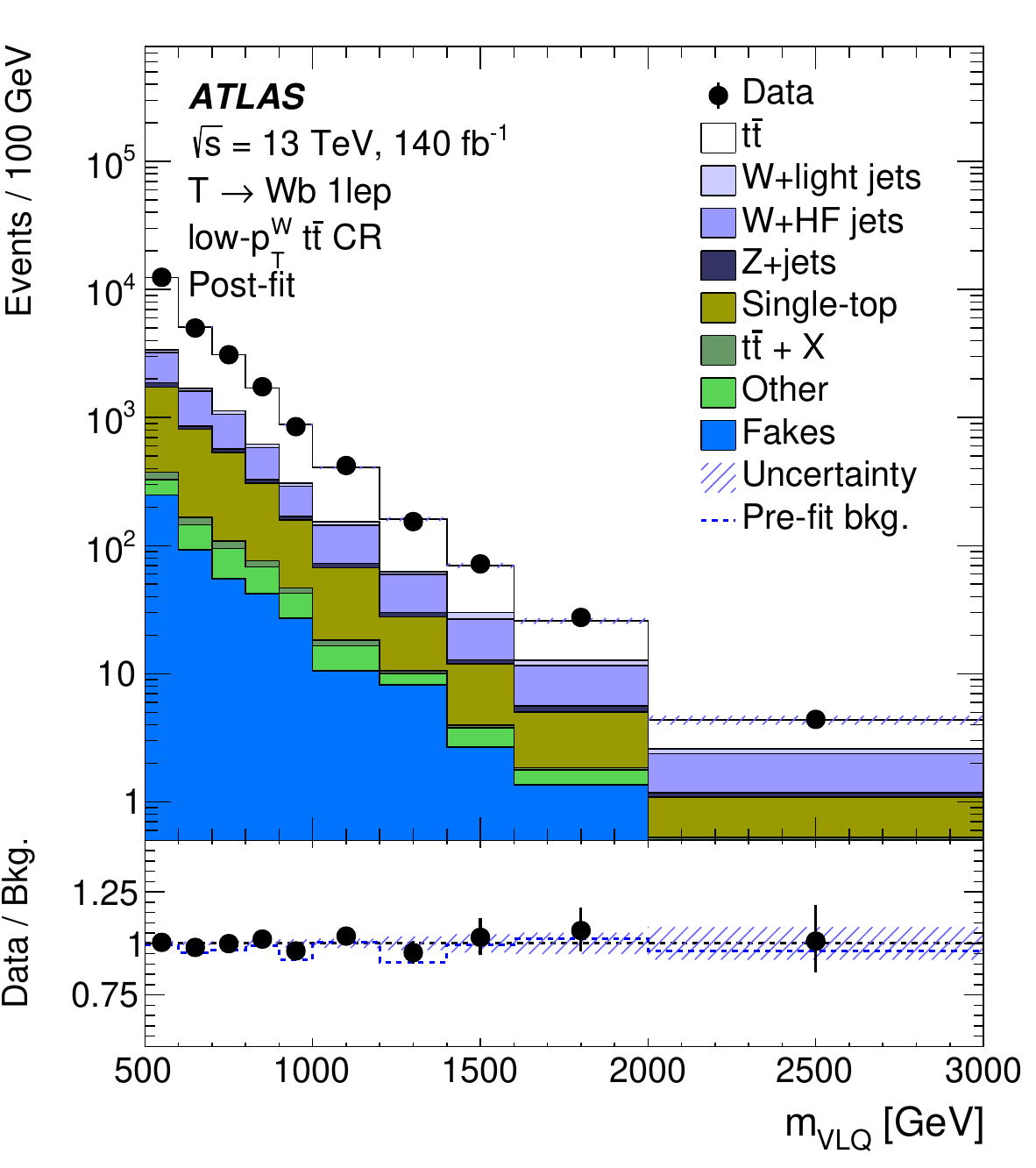}%
\label{Fig:ttcr2_lowwpt_postfit_unblinded}%
}%
\hfill
\subfigure[]{
\includegraphics[width=0.32\textwidth]{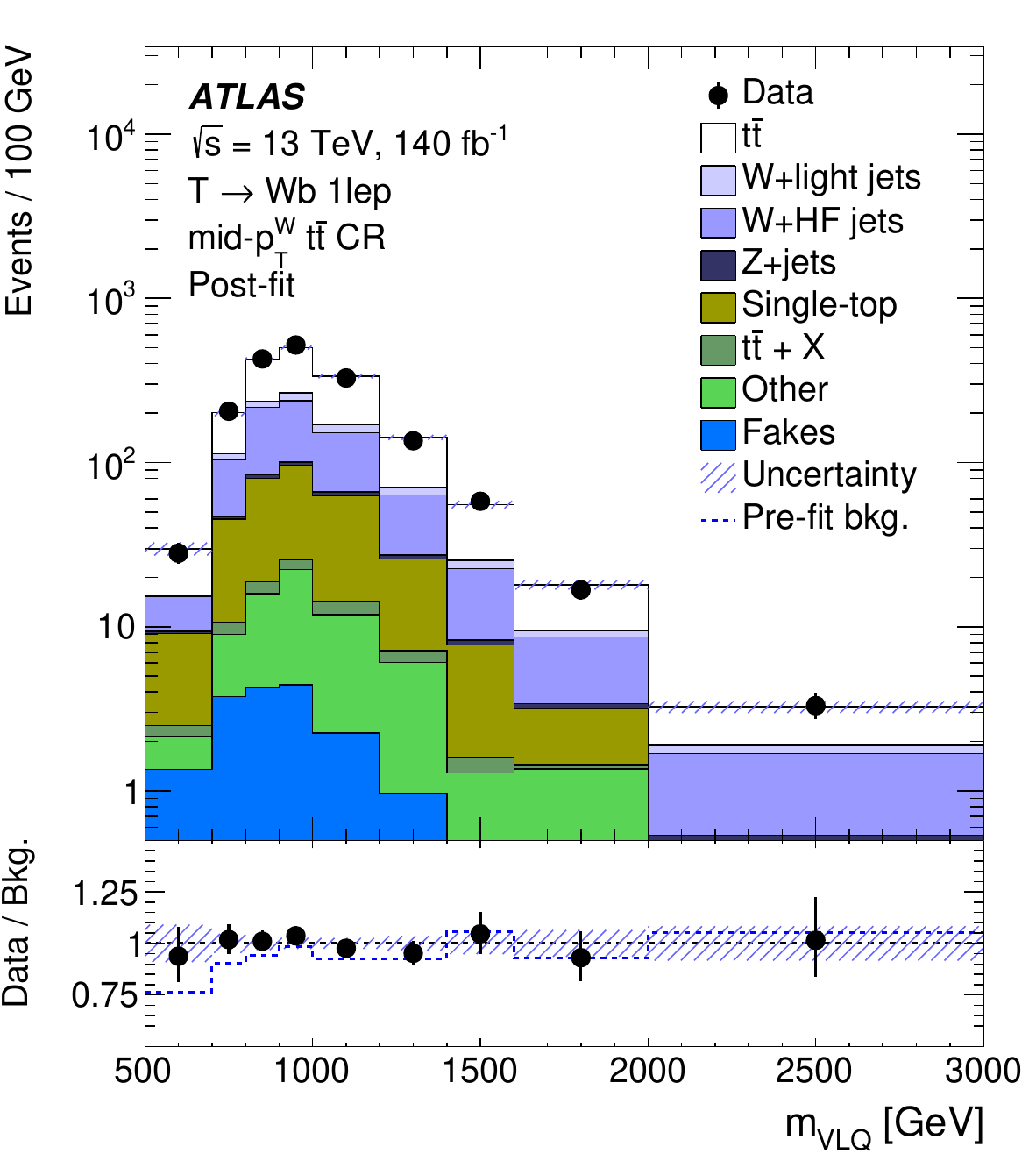}%
\label{Fig:ttcr2_midwpt_postfit_unblinded}%
}%
\hfill
\subfigure[]{
\includegraphics[width=0.32\textwidth]{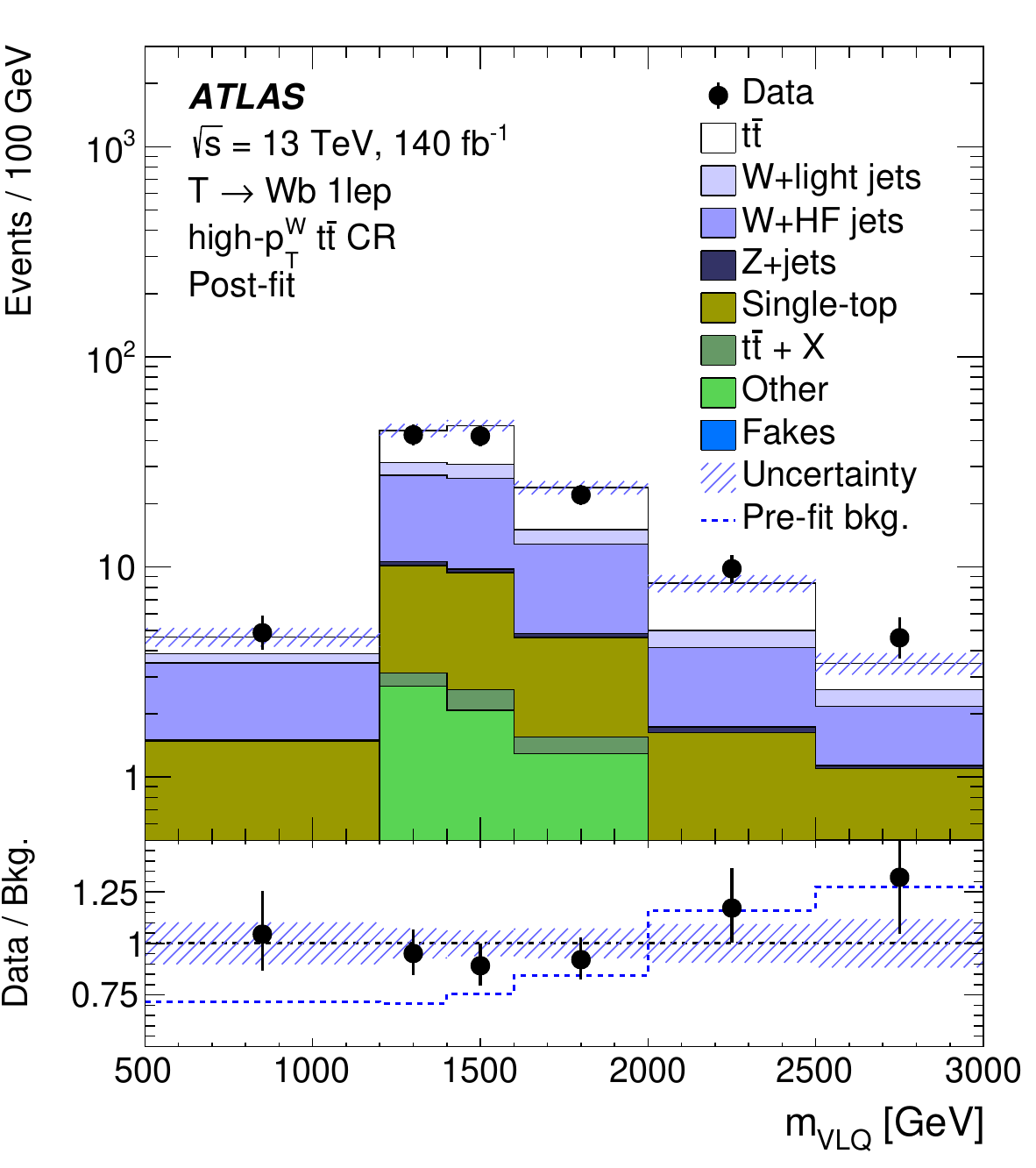}%
\label{Fig:ttcr2_highwpt_postfit_unblinded}%
}%
\caption{Distributions of the VLQ-candidate mass, $\mVLQ$, in the
(a--c) SRs, (d--f) \Wjets CRs and (g--i) \ttbar{} CRs
after the fit to the background-only hypothesis. The columns correspond
from left to right to the low-, middle-, and high-\pTW bins in each region.
\enquote{Other} includes remaining backgrounds from top quarks or that contain two $W/Z$ bosons. The last bin includes overflow.
The lower panels show the ratios of data to the fitted total
prediction. The error bars represent the statistical uncertainty
of the data. The band represents the total systematic uncertainty
after the maximum-likelihood fit.
The dashed line represents the data-to-background ratio before the fit.
The worst agreement is in the low-\pTW subregion of the \Wjets CR,
where the $\chi^2$ probabilities before and after the fit are 0.018 and 0.57, respectively.}
\label{Fig:fitresults_postfit_all_unblinded}%
\end{figure}

\begin{table}[bthp]
\begin{center}
\caption{\Wjets control region yields after the background-only fit.}
\label{tab:wj_CR_yields_postfit}
\begin{tabular}{l
S[table-format=5.2(2)]
S[table-format=5.2(2)]
S[table-format=4.3(2)]}
\toprule
& {\Wjets low-\pTW} & {\Wjets mid-\pTW} & {\Wjets high-\pTW} \\
\midrule{\ttbar}            & 9800 \pm 1400 & 2420 \pm 320 & 330 \pm 120 \\
{\WLF jets}         & 9700 \pm 1800 & 3600 \pm 700 & 1100 \pm 200 \\
{\WHF jets}         & 13900 \pm 3000 & 3700 \pm 930 & 740 \pm 230 \\
{$Z$+jets}          & 866 \pm 67 & 163 \pm 42 & 24.2 \pm 3.1 \\
{Single top}        & 1630 \pm 490 & 560 \pm 260 & 76 \pm 42 \\
{$\ttbar V$}        & 36.6 \pm 4.2 & 19.2 \pm 2.5 & 6.17 \pm 0.93 \\
{$\ttbar H$}        & 7.87 \pm 0.93 & 2.36 \pm 0.30 & 0.55 \pm 0.10 \\
{Other}             & 957 \pm 66 & 429 \pm 38 & 140 \pm 15 \\
{Fakes ($e$)}       & 320 \pm 160 & 18.5 \pm 9.2 & 17.9 \pm 8.8 \\
{Fakes ($\mu$)}     & 970 \pm 450 & 94 \pm 43 & 12.7 \pm 5.8 \\
\midrule{Total background}  & 38210 \pm 210 & 11020 \pm 110 & 2410 \pm 50 \\
\midrule{Data}              & 38191 & 11046 & 2425 \\
\bottomrule
\end{tabular}
\end{center}
\end{table}

\begin{table}[tbhp]
\begin{center}
\caption{\ttbar control region yields after the background-only fit.}
\label{tab:tt_CR_yields_postfit}
\begin{tabular}{l
S[table-format=5.1(2)]
S[table-format=5.2(2)]
S[table-format=4.3(2)]}
\toprule
& {\ttbar CR low-\pTW} & {\ttbar CR mid-\pTW} & {\ttbar CR high-\pTW} \\
\midrule{\ttbar}            & 16890 \pm 840 & 1120 \pm 140 & 122 \pm 36 \\
{\WLF jets}         & 410 \pm 120 & 117 \pm 33 & 34.7 \pm 9.8 \\
{\WHF jets}         & 3240 \pm 710 & 640 \pm 160 & 130 \pm 40 \\
{$Z$+jets}          & 274 \pm 35 & 22.0 \pm 3.2 & 3.24 \pm 0.91 \\
{Single top}        & 2940 \pm 740 & 340 \pm 130 & 56 \pm 35 \\
{$\ttbar V$}        & 67.2 \pm 7.1 & 13.4 \pm 1.8 & 3.56 \pm 0.62 \\
{$\ttbar H$}        & 30.0 \pm 3.1 & 3.59 \pm 0.43 & 0.599 \pm 0.087 \\
{Other}             & 234 \pm 22 & 72 \pm 13 & 17.7 \pm 4.8 \\
{Fakes ($e$)}       & 137 \pm 68 & 3.3 \pm 1.6 & 1.54 \pm 0.76 \\
{Fakes ($\mu$)}     & 380 \pm 170 & 21.0 \pm 9.6 & 2.2 \pm 1.0 \\
\midrule{Total background}  & 24600 \pm 180 & 2354 \pm 48 & 371 \pm 18 \\
\midrule{Data}              & 24582 & 2348 & 363 \\
\bottomrule
\end{tabular}
\end{center}
\end{table}

\begin{table}[tbhp]
\begin{center}
\caption{Signal region yields after the background-only fit.}
\label{tab:SR_yields_postfit}
\begin{tabular}{l
S[table-format=5.2(2)]
S[table-format=5.3(2)]
S[table-format=4.3(2)]}
\toprule
& {SR low-\pTW} & {SR mid-\pTW} & {SR high-\pTW} \\
\midrule{\ttbar}            & 650 \pm 120 & 345 \pm 71 & 87 \pm 36 \\
{\WLF jets}         & 443 \pm 88 & 499 \pm 98 & 150 \pm 30 \\
{\WHF jets}         & 570 \pm 130 & 550 \pm 130 & 116 \pm 36 \\
{$Z$+jets}          & 47.3 \pm 6.5 & 20.6 \pm 2.9 & 4.70 \pm 0.97 \\
{Single top}        & 371 \pm 97 & 189 \pm 53 & 23 \pm 19 \\
{$\ttbar V$}        & 5.2 \pm 1.3 & 3.76 \pm 0.94 & 1.20 \pm 0.25 \\
{$\ttbar H$}        & 1.62 \pm 0.23 & 0.768 \pm 0.096 & 0.240 \pm 0.055 \\
{Other}             & 51.1 \pm 7.7 & 63.1 \pm 8.8 & 24.7 \pm 3.6 \\
{Fakes ($e$)}       & 27 \pm 13 & 16.5 \pm 8.1 & 2.0 \pm 1.0 \\
{Fakes ($\mu$)}     & 46 \pm 21 & 11.7 \pm 5.4 & 2.5 \pm 1.1 \\
\midrule{Total background}  & 2212 \pm 49 & 1696 \pm 40 & 412 \pm 19 \\
\midrule{Data}              & 2217 & 1694 & 412 \\
\bottomrule
\end{tabular}
\end{center}
\end{table}

First, a fit to the data is performed for the background-only
hypothesis, i.e.~it depends only on $\theta$, with $\mu = 0$.
Figure~\ref{Fig:fitresults_postfit_all_unblinded} presents the \mVLQ{}
distributions after the simultaneous fit of the SR, \Wjets CR, and
\ttbar{} CR, each split into the low-, middle- and high-\pTW{}
subregions used in the fit. The fitted values of the overall
normalisation factors for \WLF{}, \WHF{} and \ttbar contributions are
found to be $0.87 \pm 0.17$, $1.09 \pm 0.24$, and $1.08 \pm 0.15$,
respectively. It is possible to extract separate factors for \WLF{}
and \WHF{} due to the \pTW{}-based splitting in both the \Wjets{} and
\ttbar CRs.  The background model and the data are in good agreement,
also in the validation regions. The
tables~\ref{tab:wj_CR_yields_postfit}--\ref{tab:SR_yields_postfit}
show the yields after the fit, including their statistical and
systematic uncertainties. Note that the uncertainty on the total
background after the fit is smaller than the quadratic sum of the
components, because of correlations in the post-fit uncertainties.

\afterpage{\clearpage}

\subsection{Limits on VLQ production}
\label{limits}
When fitting the signal models to the data, the largest observed local
signal significance is 1.5$\sigma$, corresponding to a local $p$-value
of 0.064, at a mass of 1500~\GeV. In addition, since the
background-only fit shows compatibility with the data,
profile-likelihood fits are performed with upper-limit test
statistics~\cite{Cowan:2010js},
to obtain the expected and observed 95\%~CL exclusion constraints on
the $T$/$Y$ VLQ models. The models considered all have a single
active VLQ species through a combination of multiplet structure and
chiral couplings. VLQ interference with the SM affects the \mVLQ{}
distribution in two models (the $T$-singlet, and left-handed $Y$ in a
$(T,B,Y)$ triplet); for these, interference effects
($\sigma_\mathrm{I}$) are accounted for, as described in
Section~\ref{sec:StatMethods}, although in the case of the $Y$ these
are negligible.

\begin{figure}%
\centering
\subfigure[]{
\includegraphics[width=0.45\textwidth]{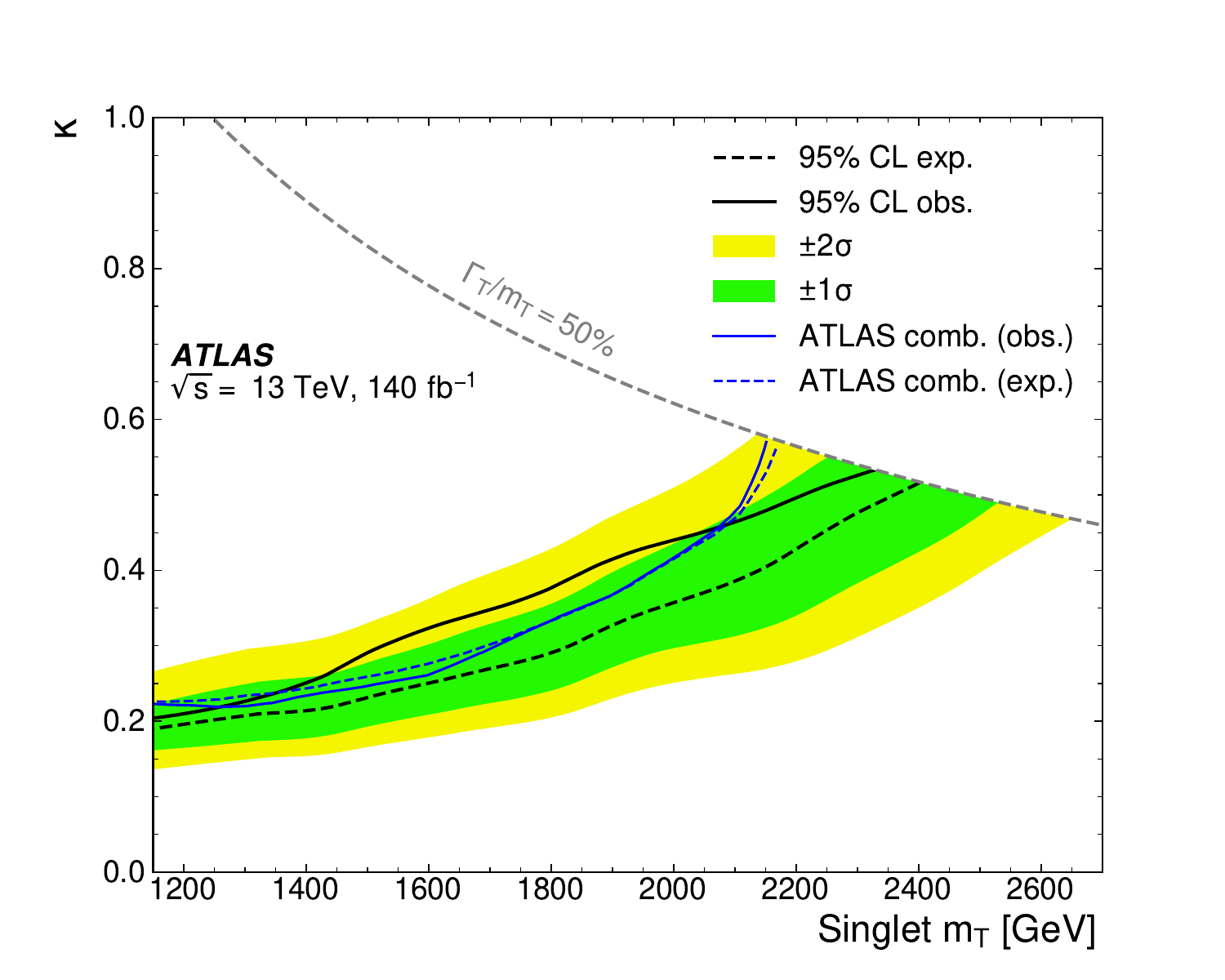}%
\label{Fig:exclusion_2D_Tkappa}%
}%
\subfigure[]{
\includegraphics[width=0.45\textwidth]{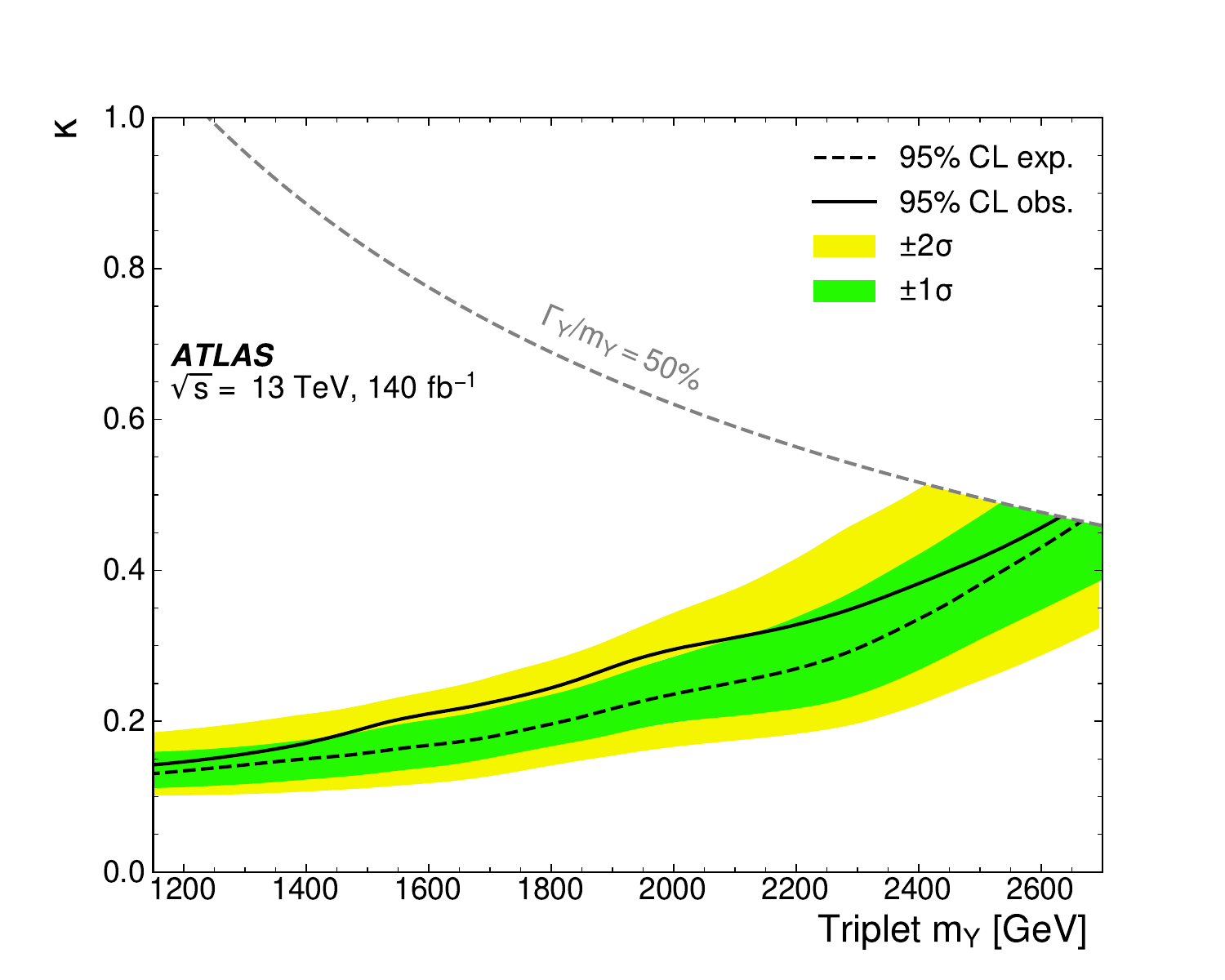}%
\label{Fig:exclusion_2D_Ykappa}%
}
\caption{Expected (background-only) and observed 95\%~CL
upper limits on the VLQ coupling $\kappa$ as functions of the VLQ
mass $\mQ$, for $T$-singlet (a) and $Y$-triplet (b)
vector-like quarks. The green and yellow bands indicate the
systematic uncertainties which are profiled in the fit for the observed upper limits.
For the $T$-singlet model, the result of the latest ATLAS combination of searches
for single vector-like top-quarks~\cite{EXOT-2021-02} is overlaid;
this includes the decay modes $T\to H t$ and $T\to Z t$.  These
interpretations are limited to parameter combinations for which
the model corrections are valid, by restricting the VLQ relative
width to $\Gamma_Q/\mQ < 0.5$, as indicated by the grey dashed
line.}
\label{Fig:exclusion_2D}%
\end{figure}

\goodbreak

Scans over \mQ{} versus $\kappa$
are presented for the $T$-singlet and $Y$-triplet in
Figure~\ref{Fig:exclusion_2D}.
As in Ref.~\cite{Deandrea:2021vje}, these interpretations are
limited to regions of model space in which $\Gamma_Q/\mQ < 0.5$, to
ensure the validity of the correction factors for the finite-width
approximation and non-resonant contributions, as noted in
Section~\ref{sec:samples}.
Upper limits at 95\% CL are set on
the single-$Q$ production cross-section times $Q \to Wb$ branching ratio
as a function of the VLQ mass for a coupling value of $\kappa = 0.5$, and
range from $40~\fb$ to $9~\fb$ between $\mVLQ$ values of $1.2~\TeV$
and $2.5~\TeV$.
The effect of the different theory cross-sections in the two VLQ
models is visible:\ in the $Y$-triplet model, with a higher theory
cross-section, a given cross-section limit is reached with a lower
$\kappa$ multiplier, resulting in stronger $\kappa$ limits.
The $T$-singlet 95\%~CL upper limit on $\kappa$ ranges from
$0.22$ at the lowest masses to $0.52$ at $\mQ \approx 2300~\GeV$;
at this point the high relative width of the VLQ reaches the edge of
validity for the NLO and finite-width corrections. Due to its higher
theory cross-section, more stringent limits are set for the $Y$-triplet,
ranging from $\kappa < 0.14$ at the lowest mass to $\kappa < 0.46$
at $\mQ = 2600~\GeV$.
The largest contributions to the uncertainty come from systematic effects
in the $\ttbar$ and single-top MC generators and the
$\ttbar$ and $W$+jets reweighting procedure, affecting the
high-\pTW{} regions especially.


%

%

%
%
\FloatBarrier

\section{Conclusion}
\label{sec:conclusion}


A search for the production of a single vector-like quark $Q$ decaying
into $Wb$, where $Q$ can be either a $T$ or $Y$ quark, has been
performed with the ATLAS experiment at the CERN LHC. The data used in
this search correspond to an integrated luminosity of
140~\ifb of $pp$ collisions with a centre-of-mass energy
$\sqrt{s} = 13~\TeV${} recorded between 2015 and 2018, and supersedes
the previous ATLAS search for this process performed with 36.1~\ifb of
data taken in 2015 and 2016.

Events with exactly one isolated electron or muon, a high-\pT{}
$b$-tagged jet, missing transverse momentum, and at least one
forward jet were selected.
A $Q$ candidate was then reconstructed and its mass used as the
discriminating variable in a maximum-likelihood fit in bins of \pTW.
The observed data distributions are compatible with the expected
Standard Model background and no significant excess is observed.

The results were interpreted for $Q=T$ in a $T$-singlet model and
$Q=Y$
in a $(T,B,Y)$ triplet, taking into account the effects of
signal-model interference with the Standard Model background.  Upper
limits at 95\% CL were set on the single-$Q$ production cross-section
times $Q \to Wb$ branching ratio as a function of the VLQ mass for a
coupling value of $\kappa = 0.5$, and limits were set on $\kappa$,
assuming the theory production cross-sections for the VLQ models.

These results complement those of a similar ATLAS search in the fully
hadronic final state, while reinforcing and extending the mass
coverage of ATLAS' single-VLQ upper limits.  The observed limits are
slightly weaker than the expectation, making the present search less
constraining than the current ATLAS single $T\to H t$ and $T\to Z t$
search combination, instead of setting stronger limits across the
considered mass range as expected. However, they will provide
important contributions to future combinations of searches sensitive
to BSM physics.


%

%
\section*{Acknowledgements}
%

%
%

%
%

We thank CERN for the very successful operation of the LHC and its injectors, as well as the support staff at
CERN and at our institutions worldwide without whom ATLAS could not be operated efficiently.

The crucial computing support from all WLCG partners is acknowledged gratefully, in particular from CERN, the ATLAS Tier-1 facilities at TRIUMF/SFU (Canada), NDGF (Denmark, Norway, Sweden), CC-IN2P3 (France), KIT/GridKA (Germany), INFN-CNAF (Italy), NL-T1 (Netherlands), PIC (Spain), RAL (UK) and BNL (USA), the Tier-2 facilities worldwide and large non-WLCG resource providers. Major contributors of computing resources are listed in Ref.~\cite{ATL-SOFT-PUB-2025-001}.

We gratefully acknowledge the support of ANPCyT, Argentina; YerPhI, Armenia; ARC, Australia; BMWFW and FWF, Austria; ANAS, Azerbaijan; CNPq and FAPESP, Brazil; NSERC, NRC and CFI, Canada; CERN; ANID, Chile; CAS, MOST and NSFC, China; Minciencias, Colombia; MEYS CR, Czech Republic; DNRF and DNSRC, Denmark; IN2P3-CNRS and CEA-DRF/IRFU, France; SRNSFG, Georgia; BMFTR, HGF and MPG, Germany; GSRI, Greece; RGC and Hong Kong SAR, China; ICHEP and Academy of Sciences and Humanities, Israel; INFN, Italy; MEXT and JSPS, Japan; CNRST, Morocco; NWO, Netherlands; RCN, Norway; MNiSW, Poland; FCT, Portugal; MNE/IFA, Romania; MSTDI, Serbia; MSSR, Slovakia; ARIS and MVZI, Slovenia; DSI/NRF, South Africa; MICIU/AEI, Spain; SRC and Wallenberg Foundation, Sweden; SERI, SNSF and Cantons of Bern and Geneva, Switzerland; NSTC, Taipei; TENMAK, T\"urkiye; STFC/UKRI, United Kingdom; DOE and NSF, United States of America.

Individual groups and members have received support from BCKDF, CANARIE, CRC and DRAC, Canada; CERN-CZ, FORTE and PRIMUS, Czech Republic; COST, ERC, ERDF, Horizon 2020, ICSC-NextGenerationEU and Marie Sk{\l}odowska-Curie Actions, European Union; Investissements d'Avenir Labex, Investissements d'Avenir Idex and ANR, France; DFG and AvH Foundation, Germany; Herakleitos, Thales and Aristeia programmes co-financed by EU-ESF and the Greek NSRF, Greece; BSF-NSF and MINERVA, Israel; NCN and NAWA, Poland; La Caixa Banking Foundation, CERCA Programme Generalitat de Catalunya and PROMETEO and GenT Programmes Generalitat Valenciana, Spain; G\"{o}ran Gustafssons Stiftelse, Sweden; The Royal Society and Leverhulme Trust, United Kingdom.

In addition, individual members wish to acknowledge support from Armenia: Yerevan Physics Institute (FAPERJ); CERN: European Organization for Nuclear Research (CERN DOCT); Chile: Agencia Nacional de Investigaci\'on y Desarrollo (FONDECYT 1230812, FONDECYT 1240864, Fondecyt 3240661, Fondecyt Regular 1240721); China: Chinese Ministry of Science and Technology (MOST-2023YFA1605700, MOST-2023YFA1609300), National Natural Science Foundation of China (NSFC - 12175119, NSFC 12275265); Czech Republic: Czech Science Foundation (GACR - 24-11373S), Ministry of Education Youth and Sports (ERC-CZ-LL2327, FORTE CZ.02.01.01/00/22\_008/0004632), PRIMUS Research Programme (PRIMUS/21/SCI/017); EU: H2020 European Research Council (ERC - 101002463); European Union: European Research Council (BARD No. 101116429, ERC - 948254, ERC 101089007), European Regional Development Fund (SMASH COFUND 101081355, SLO ERDF), Horizon 2020 Framework Programme (MUCCA - CHIST-ERA-19-XAI-00), European Union, Future Artificial Intelligence Research (FAIR-NextGenerationEU PE00000013), Horizon 2020 (EuroHPC - EHPC-DEV-2024D11-051), Italian Center for High Performance Computing, Big Data and Quantum Computing (ICSC, NextGenerationEU); France: Agence Nationale de la Recherche (ANR-21-CE31-0022, ANR-22-EDIR-0002, ANR-24-CE31-0504-01); Germany: Baden-Württemberg Stiftung (BW Stiftung-Postdoc Eliteprogramme), Deutsche Forschungsgemeinschaft (DFG - 469666862, DFG - CR 312/5-2); China: Research Grants Council (GRF); Italy: Istituto Nazionale di Fisica Nucleare (ICSC, NextGenerationEU), Ministero dell'Università e della Ricerca (NextGenEU 153D23001490006 M4C2.1.1, NextGenEU I53D23000820006 M4C2.1.1, NextGenEU I53D23001490006 M4C2.1.1); Japan: Japan Society for the Promotion of Science (JSPS KAKENHI JP22H01227, JSPS KAKENHI JP22H04944, JSPS KAKENHI JP22KK0227, JSPS KAKENHI JP23KK0245, JSPS KAKENHI JP24K23939); Norway: Research Council of Norway (RCN-314472); Poland: Ministry of Science and Higher Education (IDUB AGH, POB8, D4 no 9722), Polish National Science Centre (NCN 2021/42/E/ST2/00350, NCN OPUS 2023/51/B/ST2/02507, NCN OPUS nr 2022/47/B/ST2/03059, NCN UMO-2019/34/E/ST2/00393, UMO-2022/47/O/ST2/00148, UMO-2023/49/B/ST2/04085, UMO-2023/51/B/ST2/00920, UMO-2024/53/N/ST2/00869); Portugal: Foundation for Science and Technology (FCT); Spain: Generalitat Valenciana (Artemisa, FEDER, IDIFEDER/2018/048), Ministry of Science and Innovation (MCIN \& NextGenEU PCI2022-135018-2, MICIN \& FEDER PID2021-125273NB, RYC2019-028510-I, RYC2020-030254-I, RYC2021-031273-I, RYC2022-038164-I); Sweden: Carl Trygger Foundation (Carl Trygger Foundation CTS 22:2312), Swedish Research Council (Swedish Research Council 2023-04654, VR 2021-03651, VR 2022-03845, VR 2022-04683, VR 2023-03403, VR 2024-05451), Knut and Alice Wallenberg Foundation (KAW 2018.0458, KAW 2022.0358, KAW 2023.0366); Switzerland: Swiss National Science Foundation (SNSF - PCEFP2\_194658); United Kingdom: Leverhulme Trust (Leverhulme Trust RPG-2020-004), Royal Society (NIF-R1-231091); United States of America: U.S. Department of Energy (ECA DE-AC02-76SF00515), Neubauer Family Foundation.

%
%


%
%
%
%
%
%

%
%
%
%
%
%

%
%

%
%
%
\printbibliography
\clearpage
 
\begin{flushleft}
\hypersetup{urlcolor=black}
{\Large The ATLAS Collaboration}

\bigskip

\AtlasOrcid[0000-0002-6665-4934]{G.~Aad}$^\textrm{\scriptsize 104}$,
\AtlasOrcid[0000-0001-7616-1554]{E.~Aakvaag}$^\textrm{\scriptsize 17}$,
\AtlasOrcid[0000-0002-5888-2734]{B.~Abbott}$^\textrm{\scriptsize 123}$,
\AtlasOrcid[0000-0002-0287-5869]{S.~Abdelhameed}$^\textrm{\scriptsize 119a}$,
\AtlasOrcid[0000-0002-1002-1652]{K.~Abeling}$^\textrm{\scriptsize 55}$,
\AtlasOrcid[0000-0001-5763-2760]{N.J.~Abicht}$^\textrm{\scriptsize 49}$,
\AtlasOrcid[0000-0002-8496-9294]{S.H.~Abidi}$^\textrm{\scriptsize 30}$,
\AtlasOrcid[0009-0003-6578-220X]{M.~Aboelela}$^\textrm{\scriptsize 45}$,
\AtlasOrcid[0000-0002-9987-2292]{A.~Aboulhorma}$^\textrm{\scriptsize 36e}$,
\AtlasOrcid[0000-0001-5329-6640]{H.~Abramowicz}$^\textrm{\scriptsize 157}$
% ...
\bigskip
\\

% ...
$^{17}$Department for Physics and Technology, University of Bergen, Bergen; Norway.\\
% ...
$^{30}$Physics Department, Brookhaven National Laboratory, Upton NY; United States of America.\\
% ...
$^{36}$$^{(a)}$Facult\'e des Sciences Ain Chock, Universit\'e Hassan II de Casablanca;$^{(b)}$Facult\'{e} des Sciences, Universit\'{e} Ibn-Tofail, K\'{e}nitra;$^{(c)}$Facult\'e des Sciences Semlalia, Universit\'e Cadi Ayyad, LPHEA-Marrakech;$^{(d)}$LPMR, Facult\'e des Sciences, Universit\'e Mohamed Premier, Oujda;$^{(e)}$Facult\'e des sciences, Universit\'e Mohammed V, Rabat;$^{(f)}$Institute of Applied Physics, Mohammed VI Polytechnic University, Ben Guerir; Morocco.\\
% ...
$^{45}$Physics Department, Southern Methodist University, Dallas TX; United States of America.\\
% ...
$^{49}$Fakult\"{a}t Physik , Technische Universit{\"a}t Dortmund, Dortmund; Germany.\\
% ...
$^{55}$II. Physikalisches Institut, Georg-August-Universit\"{a}t G\"ottingen, G\"ottingen; Germany.\\
% ...
$^{104}$CPPM, Aix-Marseille Universit\'e, CNRS/IN2P3, Marseille; France.\\
% ...
$^{119}$$^{(a)}$New York University Abu Dhabi, Abu Dhabi;$^{(b)}$United Arab Emirates University, Al Ain; United Arab Emirates.\\
% ...
$^{123}$Homer L. Dodge Department of Physics and Astronomy, University of Oklahoma, Norman OK; United States of America.\\
% ...
$^{157}$Raymond and Beverly Sackler School of Physics and Astronomy, Tel Aviv University, Tel Aviv; Israel.\\
% ...
$^{a}$ Also at Affiliated with an institute formerly covered by a cooperation agreement with CERN.\\
$^{b}$ Also at An-Najah National University, Nablus; Palestine.\\
$^{c}$ Also at Borough of Manhattan Community College, City University of New York, New York NY; United States of America.\\
$^{d}$ Also at Center for Interdisciplinary Research and Innovation (CIRI-AUTH), Thessaloniki; Greece.\\
$^{e}$ Also at Centre of Physics of the Universities of Minho and Porto (CF-UM-UP); Portugal.\\
$^{f}$ Also at CERN, Geneva; Switzerland.\\
% ...

\end{flushleft}


%

%
%

%

%
%
%
%

\end{document}